\documentclass[12pt,a4paper,dipdefs]{book}
\usepackage[english, activeacute]{babel}
\usepackage{rotating}
\usepackage{epsfig}
\usepackage{amsmath}
\usepackage{mathrsfs}
\usepackage{graphics}
\usepackage{psfrag}
\usepackage{amscd}
\usepackage{leqno}
\usepackage{fancyheadings}
\usepackage{amscd}
\usepackage{indentfirst} 

\textwidth=427pt\begin{scriptsize}\end{scriptsize}
\begin{document}
\begin{center}

{\Large\sc Universidad de Granada - Instituto de F\'{\i}sica Corpuscular (UV-CSIC) }
\vspace{0.2cm}

{\large\sc Departamento de F\'{\i}sica Moderna}
\vspace{0.1cm}

{\large\sc Instituto de F\'{\i}sica Corpuscular}
\vspace{1.5cm}

\mbox{\epsfig{file=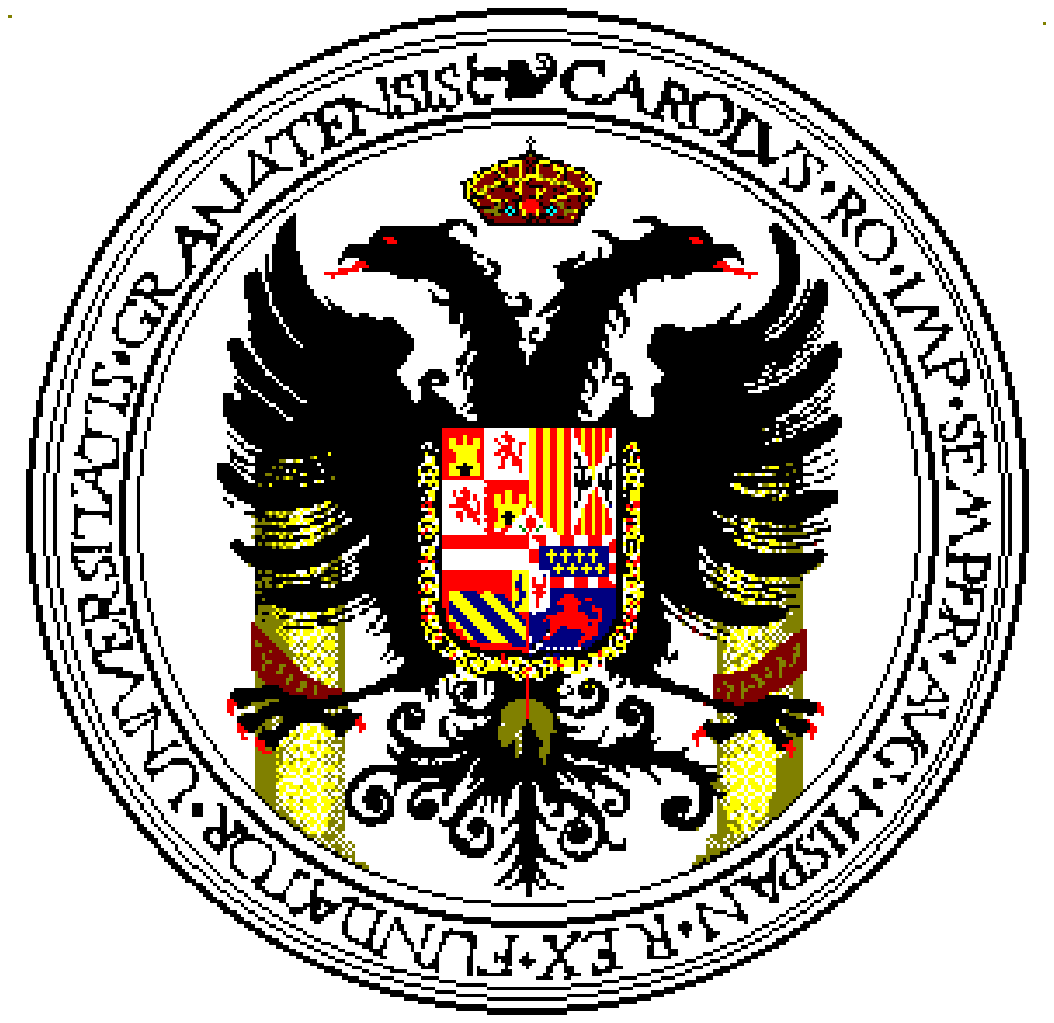,height=3.cm}}

\hspace{1.cm}

\mbox{\epsfig{file=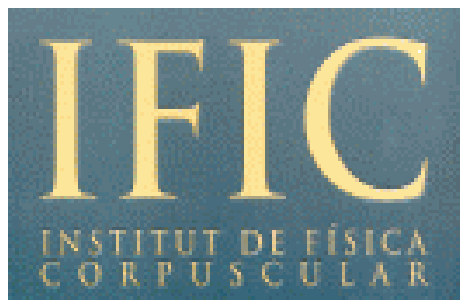,height=2.5cm}}

\vspace{1.cm}
{\LARGE\bf  Gamow-Teller Decay of the $T=1$ $^{54}$Ni  }

\vspace{4.cm}
{\large\sc Francisco Gabriel Molina Palacios} 
\vspace{0.2cm}

{\large\sc Trabajo de Investigaci\'on}
\vspace{0.2cm}

{\large\sc Julio de 2007} 
\vspace{0.2cm}

\end{center}

\vspace{4cm}

\newpage

\vspace*{4cm}

\newpage  

\begin{center}

{\Large\sc Universidad de Granada - Instituto de F\'{i}sica Corpuscular (UV-CSIC) }
\vspace{0.5cm}


{\large\sc Departamento de F\'{\i}sica Moderna}
\vspace{0.3cm}

{\large\sc Instituto de F\'{\i}sica Corpuscular}
\vspace{7cm}

{\LARGE\bf  Gamow-Teller Decay of the $T=1$ $^{54}$Ni }

\vspace{7cm}

\end{center}

 \hfill {\large\sc Francisco Gabriel Molina Palacios}
\vspace{0.3cm}

 \hfill {\large\sc Trabajo de Investigaci\'on}
\vspace{0.3cm}

 \hfill {\large\sc Julio de 2007}

\newpage

\vspace*{4.cm}

\newpage
\vspace*{4.cm}
{\bf Berta Rubio Barroso},
Investigadora Cient\'{i}fica del Consejo
Superior de Investigaciones Cient\'{\i}ficas (CSIC)
\vspace*{1.5cm}

CERTIFICA: Que la presente memoria
{\bf ``Gamow-Teller Decay of the $T=1$ $^{54}$Ni ''}
ha sido realizada bajo su direcci\'on en el Instituto de F\'{\i}sica Corpuscular
(Centro Mixto Universidad de Valencia - CSIC) por {\bf Francisco Gabriel Molina Palacios} y 
constituye su Trabajo de Investigaci\'on dentro del programa
de doctorado interunivesitario de F\'{i}sica Nuclear.
\vspace*{1.5cm}

Y para que as\'{\i} conste, en cumplimiento
con la legislaci\'on vigente,
presenta ante el Departamento de F\'{i}sica Moderna de la Facultad de Ciencias de la Universidad de Granada la
referida memoria, firmando el presente certificado en Darmstadt, Alemania a
30 de Julio de 2007.

\newpage
\vspace*{4.cm}
\newpage
\vspace*{5.cm}

\begin{flushright}
\emph{A mi Familia, que siempre han sido mis amigos y Amigos, a quienes siempre he considerado como mi familia.}
\end{flushright}

\newpage
\vspace*{4.cm}

\newpage
\pagenumbering{roman}

\tableofcontents

\listoffigures


\pagenumbering{arabic}


\pagestyle{fancyplain}


\renewcommand{\chaptermark}[1]{\markboth{\chaptername\ \thechapter}{}} 
\renewcommand{\sectionmark}[1]{\markright{\thesection\ #1}}


\lhead[\fancyplain{}{\bfseries\thepage}]{\fancyplain{}{\bfseries\rightmark}}
\rhead[\fancyplain{}{\bfseries\leftmark}]{\fancyplain{}{\bfseries\thepage}}
\cfoot{}
\thispagestyle{empty}

\chapter{Theoretical Background}
\vspace*{0.6cm}
{\it}
\section{$\beta$ Decay}
The history of beta decay began in 1896 with the discovery of radioactivity by Henri Becquerel. In the years 1899 and 1900 he identified $\beta$ radiation as one component of radioactivity, and demonstrated that $\beta$ rays are composed of electrons, comparing it with the properties of cathode rays. Unfortunately the rapidity with which $\beta$ rays were identified after their discovery did not lead to an equally rapid interpretation of the $\beta$ decay process. At that moment  physicists knew that certain substances have the same chemical properties but different radioactive properties (isotopes). As experiments improved, the interpretations of the $\beta$ decay phenomenon became more and more confusing. Chadwick in 1914, using a magnetic spectrometer, discovered the continuous spectrum of the $\beta$ particles. Thus it was known that the electron spectra had monoenergetic lines and a continuous component, in contrast with alpha and gamma ray spectra, which were known to consist of monoenergetic lines only. Chadwick further demonstrated that the majority of events were part of the continuous component, the rest being monoenergetic electrons from electron conversion. 
The interpretation of the continuous electron spectrum was the main subject of the discussion. In 1914 Rutherford thought that the $\beta$ electrons were all emitted from the nucleus with the same energy, but lost different fractions of this energy in collisions with the surrounding atoms, depending on the source thickness. But the main point was made by Lise Meitner in 1922. She realized that a quantized nucleus could not emit electrons of continuous energy. The known features of $\alpha$ and $\gamma$ spectra were correctly interpreted as due to transitions of nuclei from one quantum state to another. Thus the continuous electron spectrum was a unique feature of $\beta$ decay.\\
The measurement of the mass of different isotopes was determinant to conclude the existence of a neutral mass inside the nucleus, but the common thought was that this neutrality was due to the neutralization of the proton with an atomic electron which falls into the nucleus. Nevertheless, the uncertainty principle only allows electrons inside the nucleus with an energy greater than $\sim$100 MeV\footnote{This rough calculation was made considering the size of the nucleus $\sim$ 1 [fm], but now we know that the size of the nucleus is $\sim$ 6 [fm], and considering this value the uncertainty principle allows electrons inside the nucleus with energies $\sim$ 16 MeV} (the maximum energy of the electron $\beta$ continuous spectrum is around 10 MeV). And consequently the electrons could not exist "a priori" in the nucleus.\\

In 1927 a crucial experiment was performed by Ellis and Wooster, in which they measured the total energy released in the disintegration of a $^{210}$Bi source inside a calorimeter thick enough to stop all the emitted electrons. The endpoint of the electron $\beta$ spectrum was known to be $E_o$=1.05 MeV, and the mean energy, $\bar{E}$, of the $\beta$ electrons was known to be 390 keV. The calorimeter should have measured a total energy of 1.05 MeV if  Rutherford's reasoning  was correct. In fact they observed $\bar{E}$ = 344 $\pm$ 34 keV, which corresponded very well with the mean energy of the emitted electrons. The experiment was repeated in Berlin with an improved calorimeter by Meitner and Orthman in 1930 and the result was $\bar{E}$=337 $\pm$ 20 keV. These results were conclusive, Rutherford was wrong, but difficult to interpret at the time.\\ 
In $\beta$ decay not only was energy apparently not conserved, but the momentum and angular momentum were not conserved either.  Pauli in 1931 proposed the idea of a very penetrating neutral particle of very small mass and spin $1/2$ being emitted simultaneously with the electron. Thus, the problem of the conservation of energy, momentum and spin was solved. This proposal was made before Chadwick's discovery in 1932 of the neutral elementary particle, the neutron, and it was Fermi who named  Pauli's particle as the \emph{neutrino}. In this way the process of emission of electrons from the nucleus was explained and scientifically accepted.\\
In 1934 Irene and Fr\'ed\'eric Joliot-Curie observed the emission of positive electrons (later called positrons) in radioactive decay. And in 1938 Alvarez observed the electron capture from the nearest electronic orbits.\\
This brings us to our present knowledge of $\beta$-decay which we can describe as,
\begin{itemize}
\item[1.-] {\bf $\beta^{-}$ decay}: In this process a neutron ($n$) of the nucleus is converted into a proton ($p$) due to the week interaction, emitting an electron ($e^{-}$) and an antineutrino ($\bar{\nu}$). 
\begin{equation}
 n\rightarrow p + e^{-}+\bar{\nu}
\end{equation}
\item[2.-] {\bf $\beta^{+}$ decay}: In this process a proton ($p$) of the nucleus is converted into a neutron ($n$) due to the week interaction emitting a positron  ($e^{+}$) and a neutrino ($\nu$).
\begin{equation}
 p\rightarrow n + e^{+}+\nu
\end{equation}
\item[3.-] {\bf Electron Capture}: In the electron capture process, also called inverse $\beta^{+}$ decay, an electron ($e^{-}$) from the inner shell of the atom is captured by the nucleus, converting a proton ($p$) into a neutron ($n$) and emitting a neutrino ($\nu$).  
\begin{equation}
 p + e^{-}\rightarrow n +\nu
\end{equation}
\end{itemize}
\section{Fermi Theory of Beta Decay}
In 1934 Fermi developed a theory of the $\beta$ decay process to include the neutrino, presumed to be massless\footnote{Today we think that the neutrino has some mass, although very small} as well as chargeless, dealing with the calculation of the transition probability of the process of $\beta$ decay. This cannot be done starting from any other theory. A complete new force had to be introduced to explain the $\beta$ transition which converts a neutron into a proton (or viceversa) and at the same time produces an electron (positron) and a antineutrino (neutrino). Such a force was introduced by Fermi, using the analogy with the electromagnetism.  \\
The transition probability in the $\beta$ decay process can be given in terms of the first-order, time-dependent perturbation theory, later called the \emph{Fermi Golden Rule},
\begin{equation}
\lambda=\frac{2\pi}{\hbar}\vert\langle f \vert H_{\beta}\vert i\rangle\vert^{2}\rho_f
\end{equation}
where $\langle f \vert H_{\beta}\vert i\rangle$ is the matrix element of the beta interaction $H_{\beta}$ between the initial state $\vert i\rangle$ and the final state $\vert f\rangle$ of the complete system (nucleus and other relevant light particles), and $\rho_f$ is the density of final states in the final system. The final state of the system is specified by the electron and neutrino momenta and energies, $(p_e,E_e)$ and $(p_{\nu},E_{\nu})$, with $E_{\nu}=cp_{\nu}$. If $E_f$ denotes the energy in the final system, we have
\begin{equation}
\rho_fdE_f=V^2\frac{p^2_edp_ed\Omega_e}{(2\pi\hbar)^3}\frac{p^2_{\nu}dp_{\nu}d\Omega_{\nu}}{(2\pi\hbar)^3}
\end{equation}
where $V$ is a spherical volume in the momentum space and 
\begin{equation}
E_f=E_e+E_{\nu}=E_e+p_{\nu}c
\end{equation}
 Without fixing the directions of the momenta, we have
\begin{equation}
\rho_f= V^2(4\pi)^2\frac{p^2_edp_ep^2_{\nu}}{(2\pi\hbar)^6}\frac{dp_{\nu}}{dE_f}
\end{equation}
Fermi did not know the mathematical form of the interaction $H_{\beta}$ in $\beta$ decay. But in considering all possible forms consistent with special relativity, he showed that $H_{\beta}$ could be replaced by an operator $O_x$ which could, mathematically, take the form of a vector($V$), scalar($S$), pseudoscalar($P$), axial vector ($A$) or tensor ($T$). So, for beta decay, 
\begin{equation}\label{matrix element}
\langle f \vert H_{\beta}\vert i\rangle=W_{fi}=\sum_xg_x\int d\vec{r}\left[\Phi^*_f\psi^*_e\psi^*_{\nu}\right]O_x\Phi_i
\end{equation}
where $\Phi_f$, $\psi_e$ and $\psi_{\nu}$ are the final wave functions of the nucleus after the decay, the electron and the neutrino. The value of $g_x$ determines the \emph{strength} of the interaction. The electron and neutrino are treated like free-particles, thus their wave functions have the form,
\begin{equation}
\psi_e(\vec{r})=\frac{1}{\sqrt{V}}e^{i\vec{p_e}\cdot\vec{r}/ \hbar},\qquad \psi_{\nu}(\vec{r})=\frac{1}{\sqrt{V}}e^{i\vec{p_{\nu}}\cdot\vec{r}/ \hbar}
\end{equation}
If we expand the exponentials, using the fact that over the nuclear volume $pr\ll1$, we have the allowed approximation. If we replace the electron and neutrino wave functions in eq.\ref{matrix element} and use the allowed approximation, the matrix element is now $gM_{fi}=g\int d\vec{r}\Phi^*_fO\Phi_i$, so the decay rate is
\begin{equation}
\lambda=\frac{g^2}{2\pi^3\hbar^7}\int \frac{dp_{\nu}}{dE_f}p^2_ep^2_{\nu}\vert M_{fi}\vert^2dp_e
\end{equation}
for a fixed $E_e$ the term $\frac{dp_{\nu}}{dE_f}=1/c$. On the other hand, if we define $Q$ as the total decay energy, the momentum of the neutrino is $p_{\nu}=(Q-T_e)/c$. The decay rate is now
\begin{equation}
\lambda=\frac{g^2}{2\pi^3\hbar^7c^3}\int dp \vert M_{fi}\vert^2  p^2(Q-T_e)^2
\end{equation}
but here we are not taking into account the interaction between the beta particle and the Coulomb field in the daugther nucleus. In order to take this effect into account  an additional factor was added to the  decay rate, $F(Z',p)$ where $Z'$ is the atomic number of the daughter nucleus. The total decay rate is now
\begin{equation}
\lambda=\frac{g^2\vert M_{fi}\vert^2}{2\pi^3\hbar^7c^3}\int^{p_{max}}_0  p^2(Q-T_e)^2F(Z',p)dp.
\end{equation}
  This integral only depends on $Z'$ and the maximum electron  energy $E_0$, and it is represented as
  \begin{equation}
  f(Z',E_0)=\frac{1}{m^5_ec^7} \int^{p_{max}}_0  p^2(Q-T_e)^2F(Z',p)dp
  \end{equation}
  where the constants have been included to make $f$ dimensionless. This function is known as the \emph{Fermi integral} and it is tabulated for values of $Z'$ and $E_0$.\\
  As $\lambda=ln(2)/T_{1/2}$, we have,
  \begin{equation} \label{ft1}
  fT_{1/2}=\frac{2ln(2)\pi^3\hbar^7}{g^2m^5_ec^4\vert M_{fi}\vert^2}
  \end{equation}
  The equation \ref{ft1} gaves the comparative halflife or \emph{ft value}. 
 \section{Fermi and Gamow Teller Decay}
  
 In the allowed approximation the electron and the neutrino are created at the origin ($r=0$). As the orbital angular momentum is zero($l=0$), the only change in the angular momentum is due to the spins of the electron ($s_e=1/2$) and neutrino ($s_{\nu}=1/2$). If the two spins are antiparallel ($S=0$), there is no change in the nuclear spin: $\Delta I= \vert I_i-I_f \vert =0$. This is known as \emph{Fermi Decay} ($F$). If the electron and neutrino spins are parallel ($S=1$), they carry away an angular momentum of $1$, therefore $I_i$ and $I_f$ must be coupled through a vector of length 1: $\vec{I}_i=\vec{I}_f+\vec{1}$. This is possible for $\Delta I=0$ or $1$ (except for $I_i=0$ and $I_f=0$, when only Fermi transitions can contribute). This is known as \emph{Gamow-Teller Decay} ($GT$). They are governed by different operators which we will called $O_F$ and $O_{GT}$ in an obvious notation. \\

 The matrix elements of $F$ and $GT$ decays must be written separately. The matrix elements for Fermi Decay, following equation \ref{matrix element}, is
 \begin{equation}
M_{fi}=M_F=g_V\int d\vec{r}\Phi^*_fO_F\Phi_i
 \end{equation}
    and for the Gamow-Teller Decay 
    
 \begin{equation}
M_{fi}=M_{GT}=g_A\int d\vec{r}\Phi^*_fO_{GT}\Phi_i
 \end{equation}
  
 The Fermi operator is expressed as $O_F=\sum^{A}_{i=1}\tau_{\pm}(i)$ where $\tau_{+}$ ($\tau_{-}$) is the isospin ladder operator converting the proton(neutron) wavefunction into a neutron(proton) wavefunction. For Gamow-Teller decay the corresponding operator is $O_{GT}=\sum^{A}_{i=1}\vec{\sigma} (i)\tau_{\pm}(i)$  where $\vec{\sigma}(i)$ are the Pauli matrices, which also act on the ith nucleon. Now we can rewrite the equation \ref{ft1} in terms of the Fermi and Gamow-Teller matrix elements.
\begin{equation} \label{ft}
  fT_{1/2}=\frac{K}{\vert M_F \vert^2 + \frac{g^2_A}{g^2_V}\vert M_{GT}\vert ^2}= \frac{K}{B(F)+\frac{g^2_A}{g^2_V}B(GT)}, \quad \textrm{where} \quad K= \frac{2ln(2)\pi^3\hbar^7}{g^2_Vm^5_ec^4}
  \end{equation}
and $B(F)$ and $B(GT)$ are the dimensionless Fermi and Gamow-Teller strengths. 

\section{Isotopic Spin of the Nucleon}
The charge symmetry and charge Independence of the nuclear interaction and the almost equality of the masses of the neutron and the proton strongly suggest that they are the same particle (the nucleon) in  two different charge states. \cite{Heckman,Roy-Nigam,Bohr-Mottelson} \\ 
In analogy with the reorientation of the two degenerate states of the electron spin under the  effect of a magnetic field in the real space, a nucleon under the influence of a electromagnetic field, can differentiate the two degenerate \emph{isospin states} or \emph{charge states} in the \emph{charge space} or \emph{isospin space}. A nucleon with third isospin component $T_z=-1/2$ (down) is defined as a proton, and a nucleon with third isospin component $T_z=+1/2$ (up) is defined as a neutron.\footnote{The isospin assignment for a neutron and a proton, which is adopted here, is conventional in nuclear physics and has the advantage that the heavy nuclei with large neutron excess have their isospin aligned in the direction of the positive z axis in the isospin space. In elementary particle physics, the inverse sign convertion is usually adopted.\cite{Bohr-Mottelson}}. The isospin obeys the usual rules for angular momentum vectors. The isospin vector $\tau$ has a length of $\sqrt{\tau(\tau+1)}\hbar$ and with 3-axis projections $\tau_z=m_{\tau}\hbar$. \\ 
For a system of several nucleons, the isospin follows the same coupling rules as the ordinary angular momentum vectors. The 3-axis component of the total isospin vector, $T_z$, is the sum of the 3-axis components of the individual nucleons, and for any nucleus
\begin{center}
\begin{equation}
T_Z=\sum^{A}_{i=1}t_z(i)=\frac{1}{2}(N-Z)
\end{equation}
\end{center}
expressed in units of $\hbar$.\cite{Krane}\\

\section{Isospin Symmetry}
The assumption that the attractive nuclear force is independent of the charge of the interacting nucleons is one of the bases of our understanding of nuclear physics. This assumption hides a simple and elegant symmetry: The isospin symmetry (or charge symmetry).\\
\begin{figure}[ht!]
\begin{center}
\framebox{
\includegraphics[angle=0,width=8cm]{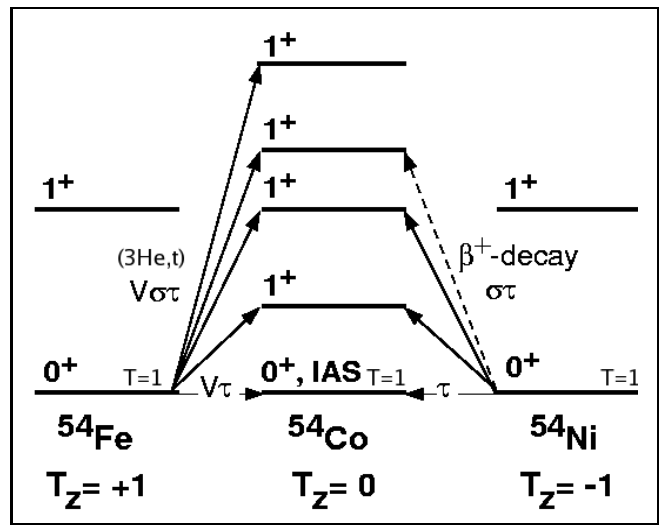}
}\end{center} 
\caption{\emph{Schematic illustration of the isospin symmetry transitions in the $A=54$ system. The level schemes, neglecting Coulomb displacement energies, are shown to illustrate the isospin symmetry structure of states in different $T_z$ nuclei.}}\label{Tequal}
\end{figure}\\
In the isospin approach, all nucleons must have an isospin $t=1/2$ differentiating protons and neutrons by its projections. The use of isospin gives us a simple way to classify nuclei according to the third component of their isospin.\\
A state in a nucleus with $N=Z$ corresponds to a particular configuration of neutrons and protons. If an identical state cannot be constructed in the neighbouring nucleus with $N-1$ neutrons and $Z+1$ protons by exchanging a neutron by a proton then the original state must have isospin $T=0$, because it can only be constructed in a nucleus with $N-Z=0$, which has projection $T_z=0$. The $N=Z$ nucleus can also have a state with $T=1$, which, because $T_z\pm1$ is also allowed, can be constructed identically in the $N\pm1$, $Z\mp1$ systems(see Figure \ref{Tequal}). \cite{WarnerNature}\\
\section{Experimental Studies of Isospin Symmetry}
The $T_z=+1$ to $T_z=0$ transitions could be studied in charge exchange (CE) reactions ($(p,n)$-type \cite{Tadd},\cite{Hagberg} or $(^3$He$,t)$-type\cite{Yoshi1},\cite{Zegers}), and the $T_z=-1$ to $T_z=0$ transitions can be investigated in $\beta$ decay experiments as shown in Figure \ref{Tequal}. Such studies have only been performed in detail so far for light nuclei, such as the $A=26$ system ($^{26}$Mg, $^{26}$Al and $^{26}$Si)\cite{Yoshi2003}. One major constraint for these studies in light nuclei is that $\beta$ decays can access only a few low-lying states due to the restriction imposed by the $Q_{\beta}$-window. Among the $T=-1\rightarrow 0$ candidates, the analogue transitions in the $fp$ shell nuclei are well suited  to an accurate study of isospin symmetry. The reason is that due to relatively large $Q_{EC}$ values, the $\beta$ decay studies should allow $B(GT)$ measurements up to high excitation energies in the daughter nuclei.  \\
\subsection{Charge Exchange reactions ($T_z=+1$ $\rightarrow$ $T_z=0$)}
It is known that the CE reactions, such as $(p,n)$ or ($^3$He,$t$) reactions, at intermediate incident energies are valuable tools for $B(GT)$ studies up to high excitation energies in the final nucleus, since they are not limited by the $Q$ window as in $\beta$ decay. It has been found that CE reactions performed at angles around $0^o$ and intermediate energies ($E_{in}>100$ MeV$/$nucleon) are good probes of $GT$ transition strength due to the approximate proportionality between the cross section at $0^o$ and the $B(GT)$ values \cite{Goodman},\cite{Tadd},
 \begin{equation}
 \frac{d}{d\Omega}\sigma_{CE}(0^o) \simeq KN_{\sigma\tau}\vert J_{\sigma\tau}(0)\vert ^2B(GT)=\hat{\sigma}_{GT}(0^o)B(GT)
 \end{equation}
where $J_{\sigma\tau}(0)$ is the volume integral of the effective interaction $V_{\sigma\tau}$ at momentum transfer $q=0$, $K$ is a kinematic factor, $N_{\sigma\tau}$ is a distortion factor, and $\hat{\sigma}_{GT}(0^o)$ is the unit cross section for a GT transition at $0^o$. CE reactions have greatly improved in recent years, because the poor energy resolution in the pioneering $(p,n)$ work has now been overcome by the use of the ($^3$He,$t$) reaction \cite{Yoshi2005},\cite{Yoshi2006}.
\subsection{$\beta$ Decay ($T_z=-1$ $\rightarrow$ $T_z=0$)}
The transition strengths $B(GT)$ are  directly obtained from $\beta$ decay \emph{ft values} which are derived from measurements of the $Q_{EC}$ value, the half-life and the branching ratio of the transition of interest. \\

In a $\beta$ decay, the partial halflife $t_{1/2}$ multiplied by the $f$-factor is related to the $B(GT)$ and the Fermi transition strength $B(F)$, as we saw in eq. \ref{ft} which has to be corrected by a Coulomb factor, 
\begin{equation}\label{ftcorrected}
fT_{1/2}=K \frac{1}{B(F)(1-\delta_c)+\lambda^2B(GT)}
\end{equation}
where $K$=6144.4$\pm$1.6, $\lambda$=$g_A/g_V$=-1.266$\pm$0.004, $\delta_c$ is the Coulomb correction factor. The Fermi strength can be calculated theoretically, and its value is $B(F)=\vert N-Z \vert $, on the contrary, the $B(GT)$ has to be taken from experiment.\\ Uncertainties in the experimental $B(GT)$ values originate from uncertainties in the decay $Q$-value, the total halflife $T_{1/2}$ and the branching ratios, determining $t$. The accurate determination of the feeding ratios to higher excited states is more difficult due to the smaller $f$-factors. On the other hand, in the $(^3$He$,t)$ measurements studying the corresponding $GT$ transition, relative transitions strength to these higher excited states can be obtained accurately. The feeding ratios in the $\beta$ decay can be deduced using these values and $f$-factors that are calculated using the decay $Q$-value. These feeding ratios can then be converted into absolute $B(GT)$ values by the normalization process using the total halflife of the $\beta$ decay.\\
The inverse of the total $\beta$-decay halflife $T_{1/2}$ is the sum of the inverse of the partial halflife $t_F$ of the Fermi transition to the Isobaric Analogue State (IAS) and the partial halflifes ($t_i$) of $GT$ transitions to the $i$th $GT$ states.
\begin{equation}
\frac{1}{T_{1/2}}= \frac{1}{t_F} + \sum_{i=GT}\frac{1}{t_i}
\end{equation}
Applying eq. \ref{ftcorrected} one can eliminate both $t_F$  and the $t_i$, so that
\begin{equation}
\frac{1}{T_{1/2}}= \frac{1}{K} \left ( B(F)(1-\delta_c)f_F + \sum_{i=GT}\lambda^2B_i(GT)f_i\right)
\end{equation}

where $f_F$ and $f_i$ are the $f$-factors of the $\beta$ decay to the IAS and to the $i$th $GT$ states, respectively, $B_i(GT)$ is the $B(GT)$ value of the transition to the $i$th $GT$ state, and $B(F)=\vert N-Z\vert$.\\
In order to relate the strengths of $GT$ and Fermi transitions in a CE reaction, the ratio $R^2$ of unit $GT$ and Fermi cross sections at 0$^o$ it is introduced.
\begin{equation}
R^2=\frac{\hat{\sigma}^{GT}}{\hat{\sigma}^F}= \frac{\sigma_i^{GT}}{B_i(GT)}\frac{B(F)(1-\delta_c)}{\sigma^F}
\end{equation}
Due to the isospin symmetry, this ratio $R^2$ is expected to be the same for all the $T_z=\pm1\rightarrow0$ transitions. Eliminating $B_i(GT)$ by using $R^2$, we get
\begin{equation}
\frac{1}{T_{1/2}}= \frac{B(F)(1-\delta_c)}{K\sigma^F}\left( \sigma^F f_F +   \frac{\lambda^2}{R^2}  \sum_{i=GT}\sigma_i^{GT}f_i \right)
\end{equation}\label{yfuji}

From this merging method point of view, $B(F)$ and $\delta_c$ depend on which nuclei  we are working with, $\sigma^F$ and $\sigma^{GT}$ are obtained from CE reactions, $f_i$ and $f_F$ are obtained from calculations and the total halflife $T_{1/2}$ is obtained from $\beta$ decay. \\

In summary, the relative values of $B_i(GT)$ can be obtained from the CE reaction. Consequently, if we have an accurate value for $T_{1/2}$ from $\beta$ decay studies, we can use the CE information to extract absolute $B_i(GT)$ values to all the states observed in the reaction. 

\chapter{Experimental Setup}
\vspace*{0.6cm}
{\it}
The experiment to measure the $^{54}$Ni beta decay was carried out at the Leuven Isotope Separator On Line (LISOL) at Centre de Recherche du Cycloron at Louvain La Neuve (Belgium)  in April 2006.

\section{The Production}

 The $^{54}$Fe target located in a gas cell was bombarded with a primary $^3$He beam at 45 MeV coming  from the CYCLONE110 cyclotron. Several radioactive nuclei were created in this reaction but since we have an on-line mass separator we are only concerned about  the production, selection and mass separation of the $^{54}$Ni and $^{54}$Co nuclei.
  
The decay of $^{54}$Ni is difficult to study because  its production is always overwhelmed by the direct production of its daugther nucleus $^{54}$Co. Indeed the calculated production cross section for the  $^{54}$Ni has a maximum of 30 $\mu$barn at an energy of 45 MeV for the $^3$He beam. At the same energy the $^{54}$Co has a production cross section of 5 $m$barn (166 times bigger)\cite{Reusentesis}. 
 
\section{The LISOL Separator}


The $^{54}$Fe($^3$He,2np)$^{54}$Co and $^{54}$Fe($^3$He,3n)$^{54}$Ni reactions are produced in the gas cell. A laser beam selectively ionizes either Ni or Co atoms. The resonant ions are extracted and accelerated  by a DC electrical field and directed to the mass separator which separates the single-charged, A=54 mass ions by $A/q$. In the following we give a detailed explanation of the different components of the LISOL device. 

\begin{figure}[ht!]
\begin{center}
\framebox{
\includegraphics[angle=0,width=12cm]{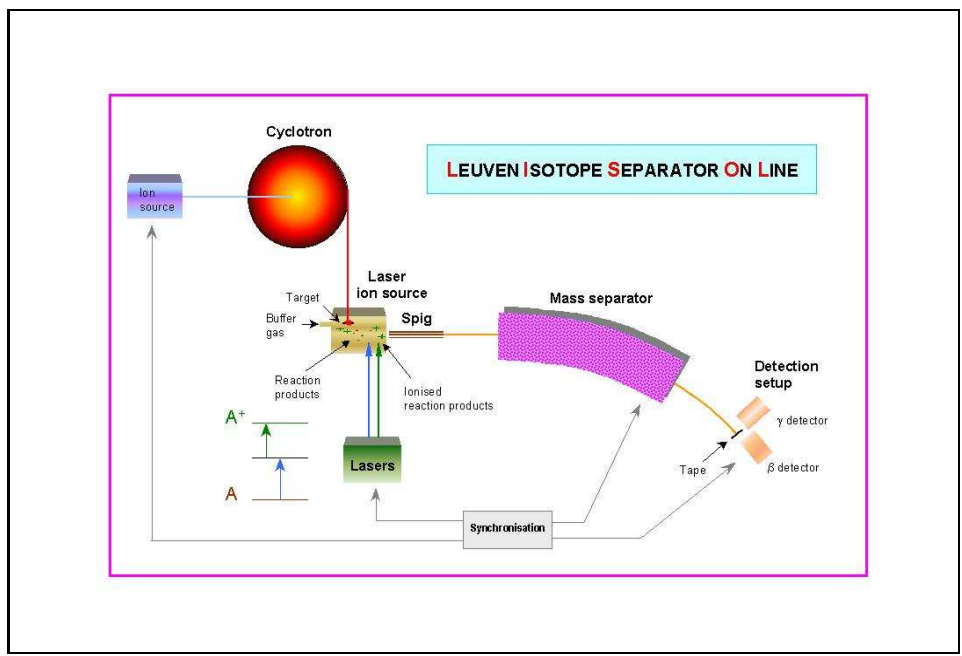}
}\end{center} 
\caption{\emph{Leuven Ion Separator On Line.} }\label{lisolfig}
\end{figure}

\subsection{Gas Cell}
The gas cell is a gas catcher source and is the first component of the so-called Laser Ion Source (LIS). The nuclear reaction occurs in the gas cell filled with a noble gas flowing through it. The reaction products are stopped, thermalized and neutralized in noble gas. This avoids chemical reactions with the nuclei produced. The body of the cell is made of stainless steel and is electro-polished to reduce the level of roughness of the surface.\\ 

In order to optimize the efficiency and the selectivity of the ion of interest, a LIS has to fulfill some important requirements such as (a) The reaction products, recoiling out the target, have to be stopped in the buffer gas, (b) all reaction products must be neutralized before arriving in the laser ionization zone, (c) avoid as much as possible the interaction of the ion of interest with impurities in the buffer gas to avoid the formation of neutral molecules, (d) the evacuation time should be shorter than the mean diffusion time to the wall in order to avoid sticking of the reaction products to the wall of the gas cell, (e) the evacuation time should be shorter than the lifetime of the isotope of interest, (f) the time between two subsequent laser pulses should be shorter than the evacuation time of the ions produced in the laser ionization region and (g) the survival time of the laser produced ions has to be longer than their evacuation time. \\

\subsection{Resonant Laser Ionization}

After the thermalization and neutralization of the reaction products we need to select and extract the ion of interest. The resonant laser ionization technique is used to ionize a specific element. \\

\begin{figure}[ht!]
\begin{center}
\framebox{
\includegraphics[angle=0,width=12cm]{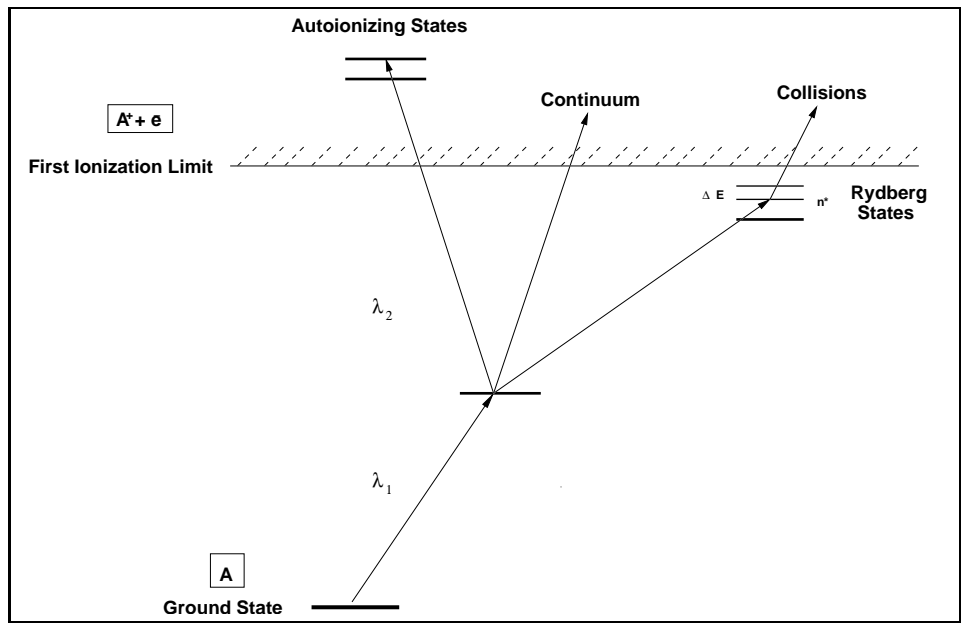}
}\end{center} 
\caption{\emph{Atomic level scheme and the different ways to ionize an atom.}}\label{laser1}
\end{figure}

Two tunable dye lasers are pumped by two excimer Xe-Cl (308 nm) lasers. The atoms thermalized in the ground state level are excited in two steps. The first step laser $\lambda_1$ excites the atom to an intermediate level. The second step laser $\lambda_2$ may ionize and excite atom to the continuum, to an autoionizing state or to a Rydberg state and where it is ionized by collision with the buffer gas (See Figure\ref{laser1}).\\
  This two dye laser beams are focused by lenses and directed by prisms to the gas cell located at 15 m from the optical setup. In order to focus the two beams at an optical distance of 15 m a screen is located at that distance but not in the same position as the gas cell.\\
To choose the most efficient resonance wavelength of the two dye laser beams, a fraction of each beam is deflected and directed to a reference cell in a  vacuum chamber. An atomic beam is produced by evaporating the element investigated from a resistively-heated crucible. \\
After the laser ionization, the ionized atoms and compound molecules\cite{Yurispig2001} are attracted by a potential difference and ejected from the gas cell.
\subsection{Sextupole Ion Guide}

Once the selected atom is ionized it can re-combine with the impurities (normally H$_2$O) in  the gas cell forming molecules\cite{Yurispig2001}. These compound molecules are heavier than a single atom so the isotope of interest is lost in the subsequent mass separation step. This problem can be solved if a sextupole ion guide is located at the exit hole of the gas cell\cite{Pietspig1997}.\\

\begin{figure}[ht!]
\begin{center}
\framebox{
\includegraphics[angle=0,width=12cm]{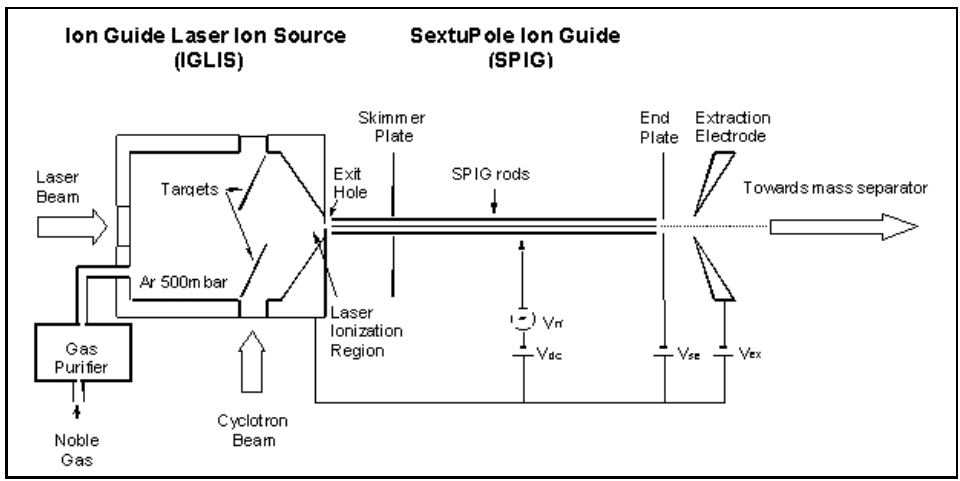}
}\end{center} 
\caption{\emph{Sextupole Ion Guide (SPIG)} }\label{spigfig}
\end{figure}
 
 The {\bf S}extu{\bf P}ole {\bf I}on {\bf G}uide SPIG, as shown in  Fig.(\ref{spigfig}) consists of six silver rods (1.5 mm diameter and 124 mm lenght) cylindrically mounted and forming a hexagonal structure parallel to the beam path. An oscillating radio frequency voltage (V$_{rf}$) is applied to the rods, with every rod in antiphase with the neighbouring rods. The amplitude and the frequency of the $rf$ signal are chosen in order to confine the ions in the centre of the SPIG. Hence the gas coming out of the gas cell can expand through the gaps between the rods, while the ions are radially confined and transported to an extraction electrode with the velocity of the gas jet.\\
 In addition to the $rf$ voltage, a $DC$ voltage is applied between the rods and the gas cell exit (grounded).
 Hence a molecule could dissociate and liberate the ion of interest, thus increasing the selection production efficiency of the LIS. 
     
 After that, the selected ion is mass-separated in a mass separator. The mass separator is  a 55$^o$ dipole magnet of 1.5 m radius. The ions are separated according to the mass-over-charge ratio since they are, in general, in a single charge state, normally $1^+$. Hence they are separated in mass. The ion beam of the mass selected is implanted into a tape located in the middle of two Miniball germanium detectors (see Figure \ref{setupexp}), as described in the next chapter. 
 \section{Radiation Detectors}
 
\begin{figure}[ht!]
\begin{center}
\framebox{
\includegraphics[angle=0,width=12cm]{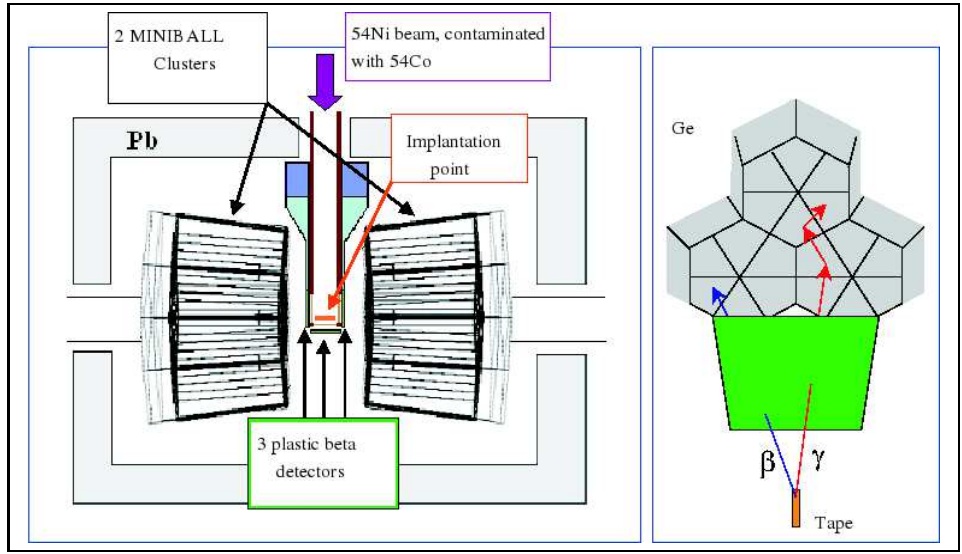}\label{setupexp}
}\end{center} 
\caption{\emph{Top and side view of the setup of $\beta$ and $\gamma$ detectors at Louvain la Neuve facilities. } }
\end{figure}

 In the $\beta$ decay of $^{54}$Ni we expect to have three type of radiation: betas, neutrinos and gammas. It is difficult to detect the neutrinos but we can make a sensitive setup to detect $\beta$ radiation and $\gamma$-rays. For the $\gamma$-ray detection we used two MINIBALL Ge cluster detectors and for the $\beta$ particles we used plastic detectors. Three $1.3$ mm thick plastic $\beta$ detectors were placed around  the implantation point (tape ending, see Figure \ref{setupexp}). In  most of the cases the $\beta$ particle only deposits  part of its energy in the thin plastic. These detectors were used to determine if a $\beta$ particle passed through the detector or not. No energy was registered. \\

\begin{figure}[ht!]
\begin{center}
\framebox{
\includegraphics[angle=0,width=12cm]{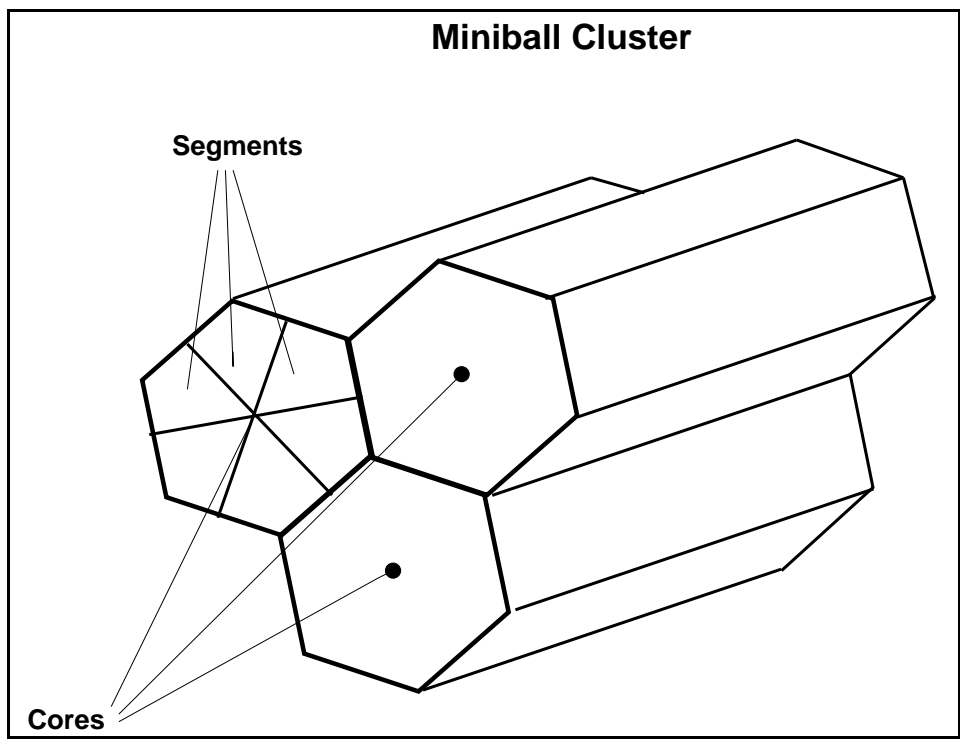}
}\end{center} 
\caption{\emph{Schematics showing cores and segments in a Miniball Cluster } }\label{ge_det}
\end{figure}

Two MINIBALL clusters were located as close as possible to two of the scintillators. The MINIBALL detector is a cluster of three hyper-pure germanium crystals, each of them six-fold segmented. The crystals and the capsules are tapered in a hexagonal shape to fit three of them side-by-side in a single cryostat. A positive high voltage is applied to a lithium-diffused centre contact, which penetrates deep inside  the germanium. Six  boron-implanted contacts on the surface of each crystal constitute the segmented outer contacts. An incident $\gamma$-ray interacts with the Ge crystal and transfers its energy either totally (photoelectric effect) or partially (Compton or pair creation) to an electron, this electron loses its kinetic energy, creating a number of electron-hole pairs. Within the fully depleted region of germanium between the contacts, the  charge created goes to the corresponding electrode. Thus, the electrons go to the centre contact (core signal), and the holes go to the outer contacts (segment signal) (See Figure \ref{ge_det}) . \\ 
The charge collected on the crystal contacts is integrated by charge-sensitive resistor feedback preamplifiers, developed and produced by the Institute of Nuclear Physics in Cologne, a member institute of the MINIBALL  Collaboration.\cite{Olegtesis} 

\section{The Tape System}
\begin{figure}[ht!]
\begin{center}
\framebox{

\includegraphics[angle=0,width=12cm]{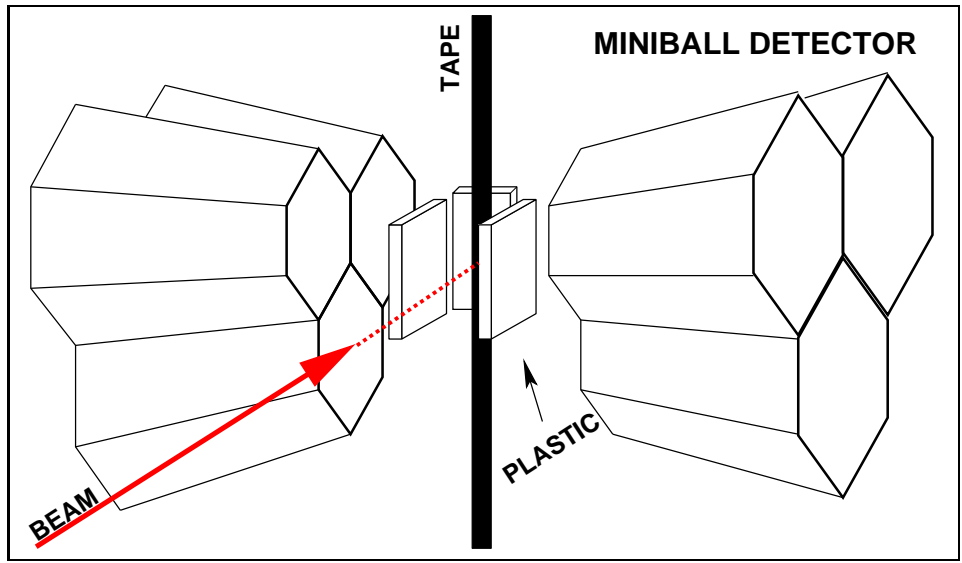}

}\end{center} 
\caption{\emph{Schematic view of the detector and type system setup } }\label{Det_setup}
\end{figure}

After the mass separation, the ions are implanted on a thin 1.25 cm wide aluminium-coated mylar tape. The thickness of the tape is chosen to stop the beam completely. The tape rewinding system is located below the detectors. The tape  from one of the two coils goes up, intercepts the beam path, goes up again and turning fully down, comes again below the detector level and finally ends in the second coil. The part of the tape where the ions are implanted is moved down to a safe distance away from the detectors, to avoid the background produced by the activity of  long-lived ions close to stability. The whole tape system is kept in the same vacuum as the beam transport line. There are three thin mylar windows in the chamber around the implantation area to reduce the attenuation of the radiation of interest.\\
A schematic picture of the detector setup and the tape system is shown in Figure \ref{Det_setup}.
 
\section{The Data Acquisition System}

The preamplified signals are provided to the analogue inputs of the DGF-4C modules. The DGF (Digital Gamma Finder) can measure the amplitude and the shape of the incoming pulses simultaneously. The energy information is extracted directly from the preamplified signal height and the time information is obtained from a built-in high-precision internal clock after the event validation.\\
In our detector system we have 21 signals from each MINIBALL $\gamma$ detector (3 signals from the cores and 18 segments) and another 3 signals from the three plastic scintillator $\beta$ detectors. Since an incident $\gamma$ ray is registered by the core and some of the segments of the same crystal, there is a time correlation between the signals from one crystal. Further on there should be no segment signal  unless there is a signal from the corresponding crystal core. For these reasons, the DGF modules are split into trigger groups, in which each group represents one crystal and the core channel is the main trigger of the group. The logic signals from the $\beta$ detectors are provided to one DFG module and their inputs are set independently. To estimate the total acquisition live time, a pulser signal feeds one channel of a DGF module.  This value is represented by the ratio between the registered pulser signals to the number of pulses provided by the pulser.\\
\section{Data Analysis}
A new C++ software code has been created by Oleg Ivanov and Dieter Pauwels from Katholieke Universiteit Leuven, to read and analyze the data acquired by the detection setup. The object oriented ROOT package developed at CERN is used for building the output histograms. While building the singles spectra is a simple procedure, the $\beta$-n$\gamma$ and n$\gamma$ coincidences, where n represents the number of $\gamma$-detectors (crystals), are based on the event time stamps. These coincidences are stored in ROOT trees for further analysis\cite{Olegtesis}.

\chapter{The Experiment}

The aim of this experiment was to study the GT transitions in the $\beta$-decay of $^{54}$Ni analogous to the GT transitions of $^{54}$Fe($^3$He,t). The main objectives are to measure 1) the total halflife $T_{1/2}$ of $^{54}$Ni with high accuracy, 2) the decay branching ratios of the $^{54}$Ni to the ground state and to the first GT state and, if  possible 3) the decay to at least one extra GT excited state in $^{54}$Co. \\
 
\begin{figure}[ht!]
\begin{center}
\framebox{
\includegraphics[angle=0,width=\textwidth]{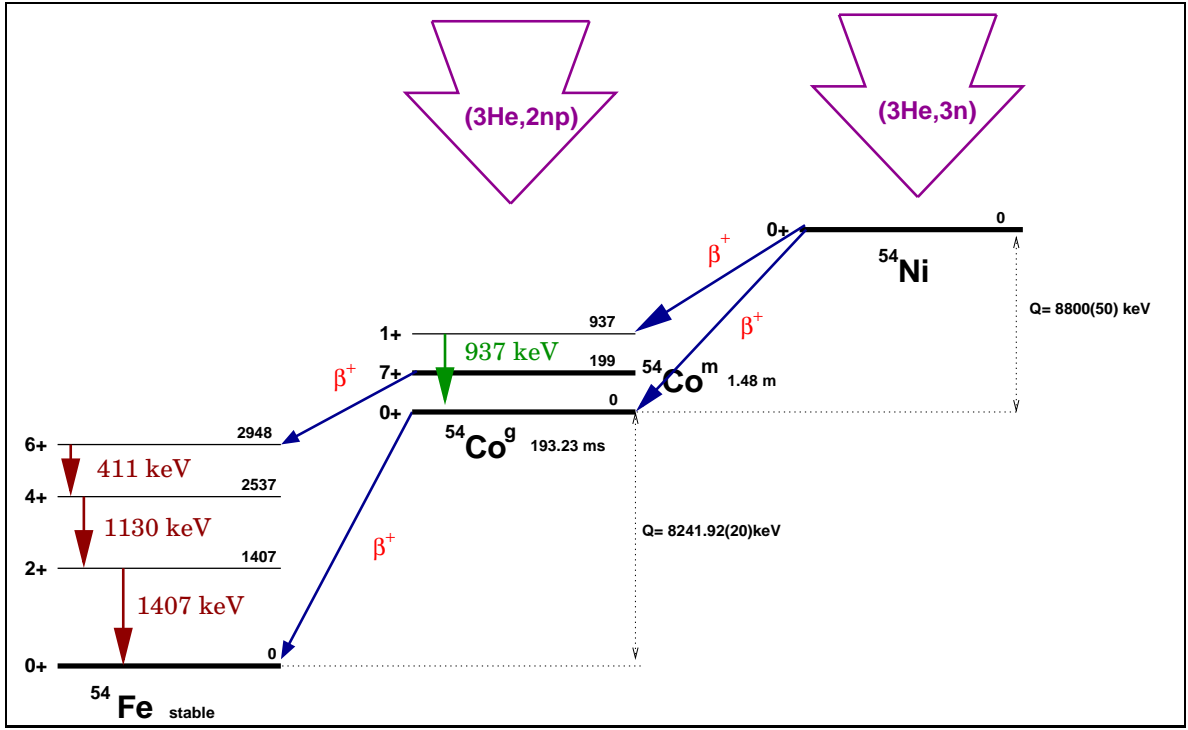}
}\end{center} 
\caption{\emph{$^{54}$Ni and $^{54}$Co$^{g,m}$ decay scheme.}}\label{level}
\end{figure}

Figure \ref{level} shows the known decay scheme of $^{54}$Ni and  $^{54}$Co$^{g,m}$. The $\beta^+$ decay of $^{54}$Ni feeds the ground state of $^{54}$Co and at least one exited state (I$^{\Pi}$=1$^{+}$) with an energy of 937 keV. The decay of the $^{54}$Co$^m$ is accompanied by a triple $\gamma$ cascade of 411 keV, 1130 keV and 1408 keV $\gamma$ rays. The $^{54}$Co ground state decays 100\% to the ground state of $^{54}$Fe.   

\section{Calibrations}
The resolution check for the MINIBALL detectors was performed with a $^{60}$Co source. The measured resolutions for all MINIBALL crystals at 1332 keV  are shown in the next table for the "Heidelberg"  and "Leuven" clusters.

 \begin{center}
\begin{tabular}{|c|c|c|c|}
\hline
Heidl.A & 3.4 keV & LeuvenA & 3.3 keV\\
\hline
Heidl.B & 2.7 keV & LeuvenB & 3.1 keV\\
\hline
Heidl.C & 2.8 keV & LeuvenC & 3.0 keV\\
\hline
\end{tabular}
\end{center}

\begin{figure}[ht!]
\begin{center}
\framebox{
\includegraphics[angle=0,width=\textwidth]{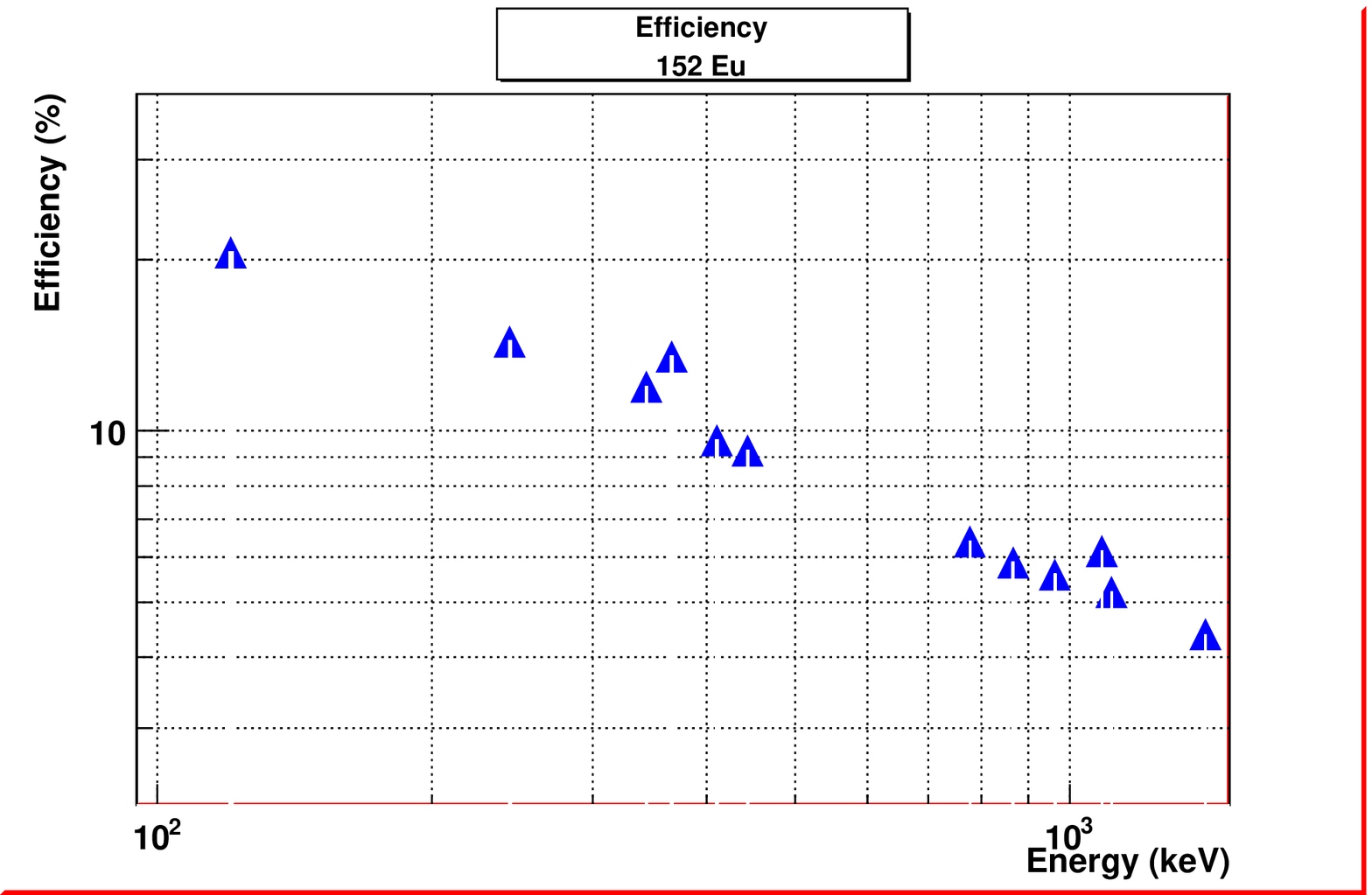}
}\end{center} 
\caption{\emph{$\gamma$ Efficiency measurement of the MINIBALL detectors with a 152Eu source after the experiment.}}\label{eff152Eu}
\end{figure}

 The $\gamma$ efficiency calibrations (See Figure \ref{eff152Eu}) were performed after the experiment with a $^{152}$Eu positioned on the tape at the point of implantation. The activity of the source at the time of the experiment  was 23.87 kBq. \\

 
The same  $^{152}$Eu source was used before the experiment, for the on-line analysis energy calibration. Afterwards, for off-line analysis, an internal energy calibration was performed using the $^{54}$Co$^m$ gamma cascade present in our experiment for each ROOT file created after more or less each one hour of running time.
\section{Control Measurements during the Experiment}
With all the calibrated spectra, several control measurements were performed before and during the experiment in order to optimize the production and mass separation of $^{54}$Ni and $^{54}$Co$^{g,m}$. 
\begin{itemize}
\item[1.-] {\bf Laser tuning on Co}: To optimize the selectivity of Co ions by laser resonance  in the gas cell, we looked at the ratio in intensity of 411 keV and 511 keV $\gamma$-ray lines from the decay of $^{54}$Co$^{g,m}$ in the singles and $\beta-\gamma$ coincidence spectra with a fixed mass 54 in the mass separator. The cyclotron current and the tape steps were also fixed. Short measurements of 100 seconds with and without laser resonance were made. 
\item[2.-] {\bf Mass tuning}: Once the selectivity was optimized, we proceeded to tune the mass selection on the separator with laser resonance on Co looking at the integral of  the 411 keV line in the singles and $\beta-\gamma$ spectra, taking short measurements of 100 seconds. 
\item[3.-]{\bf Gas Buffer Pressure and Cyclotron Current}: With the mass optimized and with laser resonance on Co, the optimal gas buffer pressure was found looking again at the 411 keV line in the singles and $\beta-\gamma$ spectra, taking short measurements of 100 seconds. After this the cyclotron current was tuned following the same procedure as described above.
\item[4.-]{\bf Tape Step and Implantation-Decay Macrocycle}: In the previous $^{54}$Ni $\beta$ decay experiment developed by I.Reusen et al.\cite{Reusen},\cite{Reusentesis} the measured halflife was 106(12) ms. Consequently, a proper 400 ms and 600 ms implantation-decay macrocycle was chosen for this experiment. The tape was moved one step every six macrocycles in order to minimize the time lost during the tape rewinding. 
\end{itemize}
In the following table the starting settings are summarized. 

\begin{center}
\begin{tabular}{|c|c|}
\hline
Mass & 53.94 \\
\hline
Cyclotron Current & 4.0 $\mu$A\\
\hline
Gas Buffer Pressure & 760 mbar\\
\hline
ID Macrocycle & 400 ms / 600 ms\\ 
\hline
Tape Movement Period & 6 Macrocycles\\
\hline
\end{tabular}
\end{center}

These settings were checked and, if  necessary, modified during the experiment in order to increase the selectivity and mass separation of the ion of interest. 
   
\section{Data Analysis}

\begin{figure}[ht!]
\begin{center}
\framebox{
\includegraphics[angle=0,width=\textwidth]{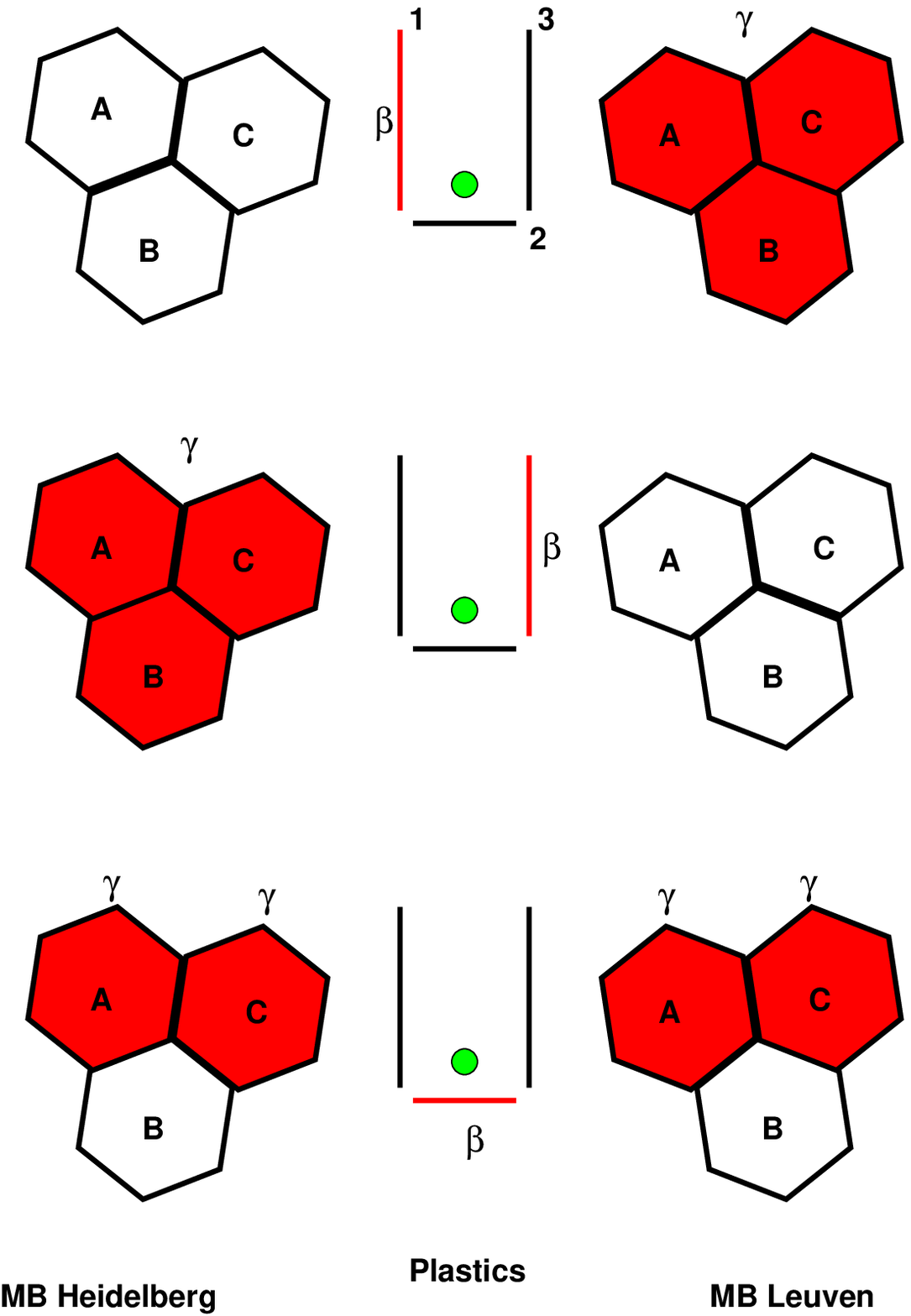}
}\end{center} 
\caption{\emph{Veto Condition: If a $\beta$ particle is detected in one of the side plastic detectors, the corresponding $\gamma$'s are accepted if they happens in the opposed MINIBALL cluster. If the $\beta$ particle goes through the middle $\beta$ detector, the $\gamma$ rays are accepted if they happens in the two most distant crystals of each MINIBALL cluster }}\label{vetocond}
\end{figure}

\begin{figure}
\centering
\fbox{
\includegraphics[angle=0,width=\textwidth]{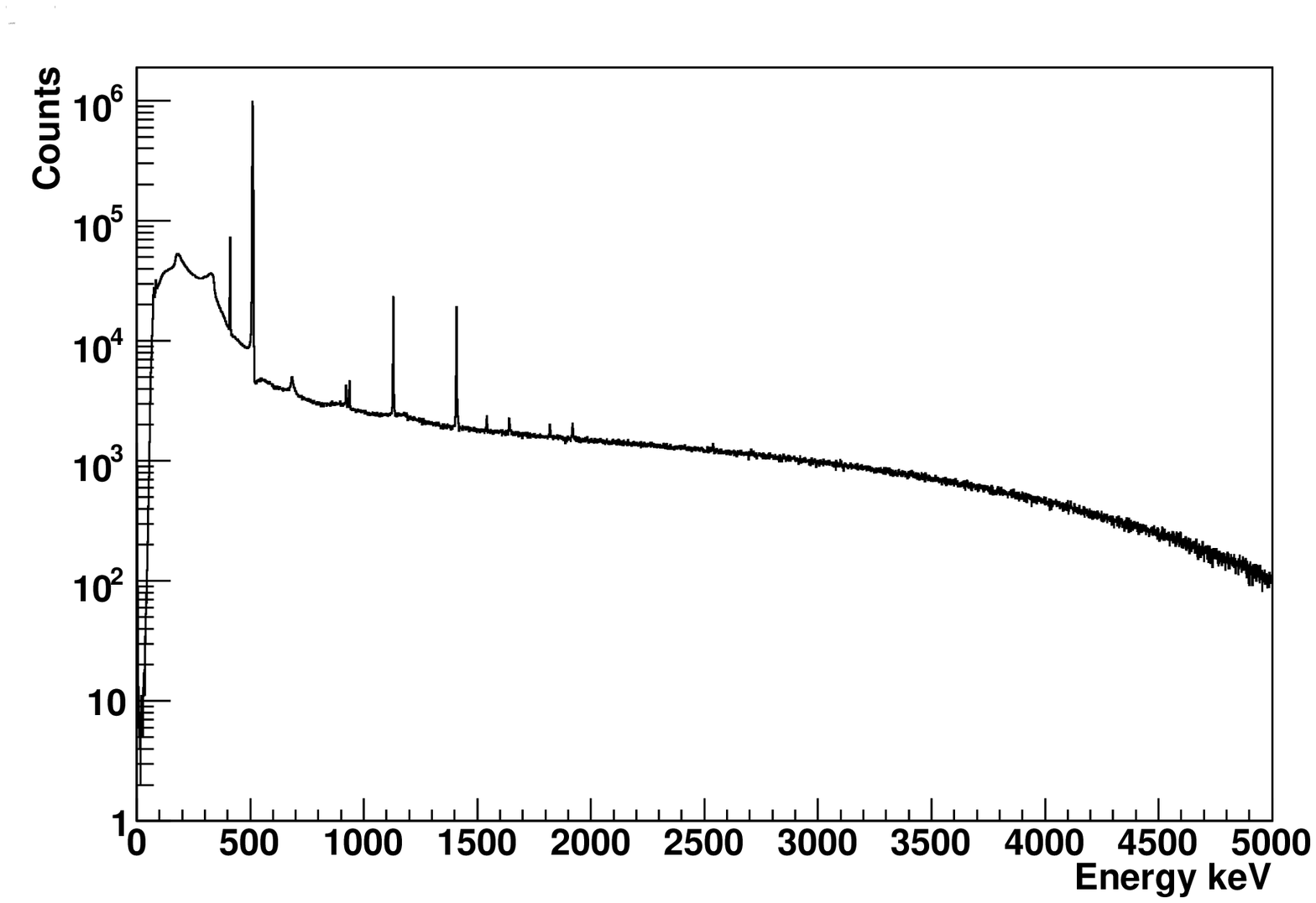}
}
\hspace{3cm}%
\fbox{
\includegraphics[angle=0,width=\textwidth]{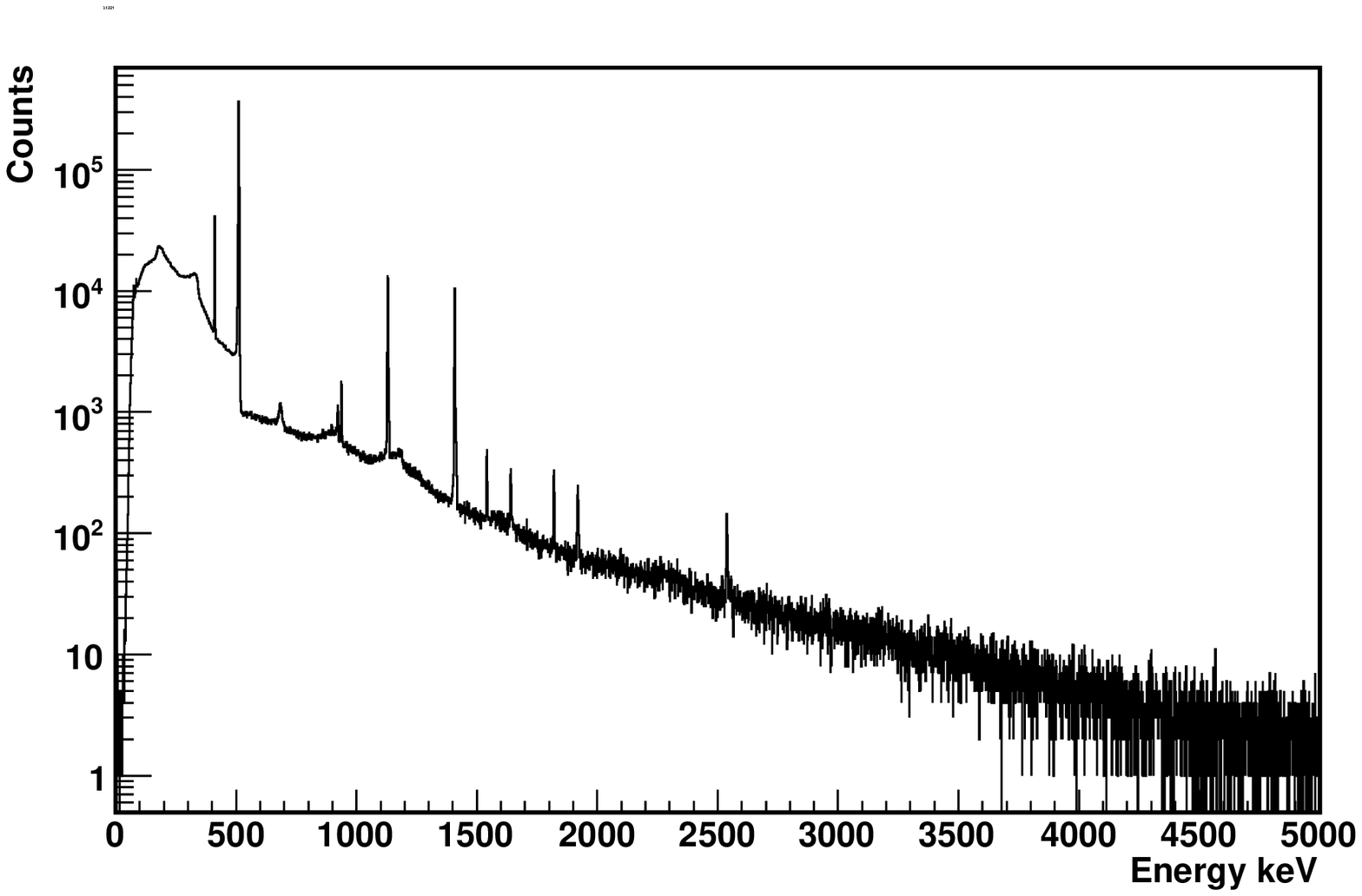}
}
\caption{\emph{Difference between the no-vetoed (up)  and vetoed (down) laser on Ni, mass A=54  $\beta$ gated $\gamma$  spectra}}\label{Veto}
\end{figure}

After seven days of experiment, the data acquisition time with laser resonance on Ni was 115 hours. The time used for the laser resonance on Co was 16 hours. The total number of counts in the 411 keV $^{54}$Co$^m$ decay radiation in the laser on Ni total spectrum was 94,817 while the number of counts in the 411 keV peak in the laser on Co total spectrum was 46,903. Finally, the settings adjustment time was 2.5 hours. Singles and $\beta$-$\gamma$ coincidence spectra were constructed in ROOT trees to facilitate the on line analysis. These pre-processed data were also used for the off-line analysis.\\
In general the data analysis was done with the $\beta$-$\gamma$ coincidence spectra. As we are interested also in the high energy GT excited states, a veto condition (see Figure \ref{vetocond}) was imposed in order to reduce the background produced by the penetration of the $\beta$ particle into the MINIBALL crystal. The main idea is that if a $\beta$ particle is detected in a plastic detector, the gamma detector situated behind it is not used.  In  Figure \ref{Veto} we see the difference between the spectrum with the no-veto condition, which has 5,984 counts in the 937 keV peak  and with the veto condition, which has 3,886 counts in the 937 keV peak. Although the upper spectrum has higher statistics, the lower one has a better peak-to-total ratio.

\begin{figure}[ht!]
\begin{center}
\framebox{
\includegraphics[angle=90,width=\textwidth]{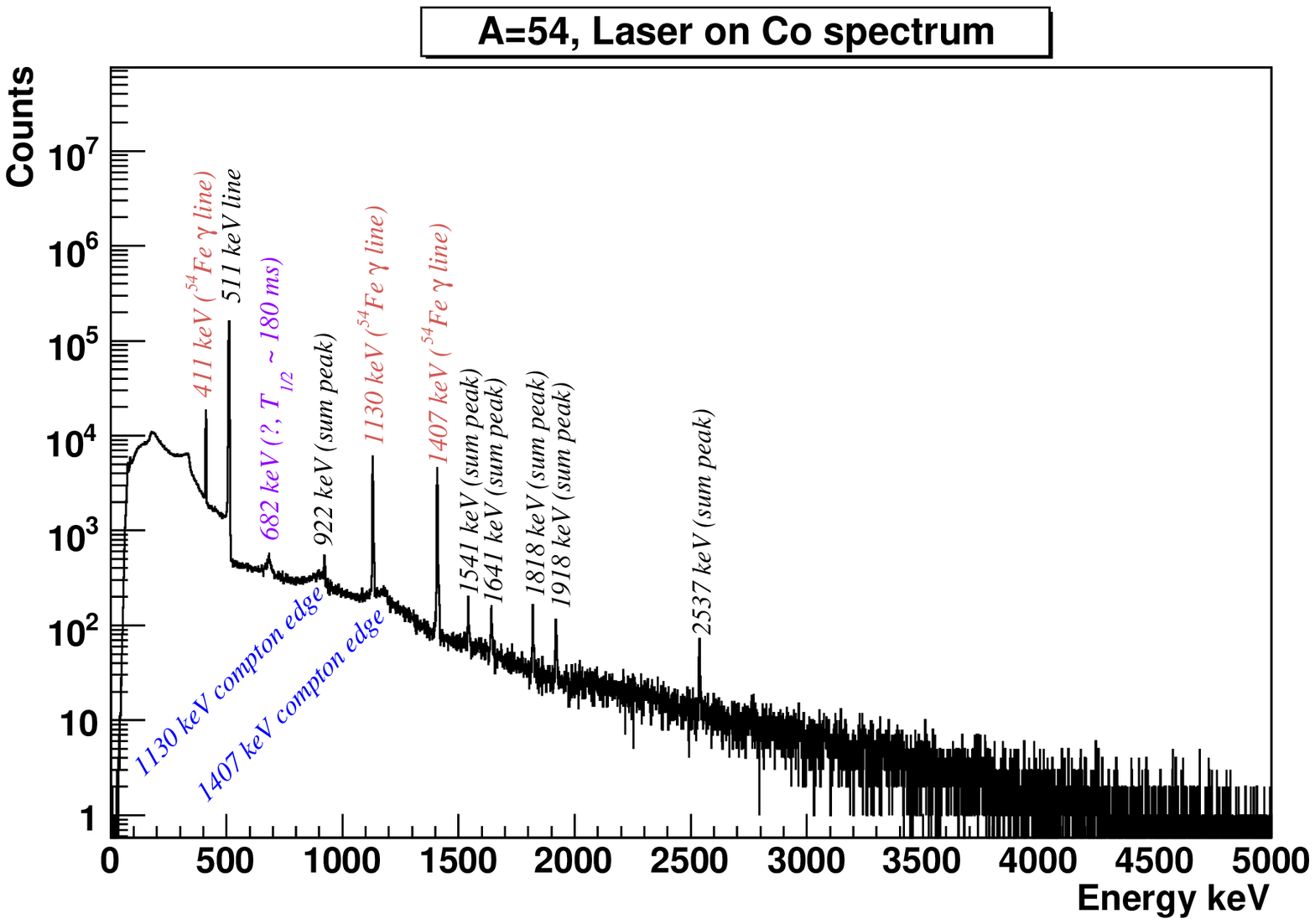}
}\end{center} 
\caption{\emph{A=54 laser on Co $\beta$-$\gamma$ spectrum}}\label{54Cototal}
\end{figure}

\begin{figure}[ht!]
\begin{center}
\framebox{
\includegraphics[angle=90,width=\textwidth]{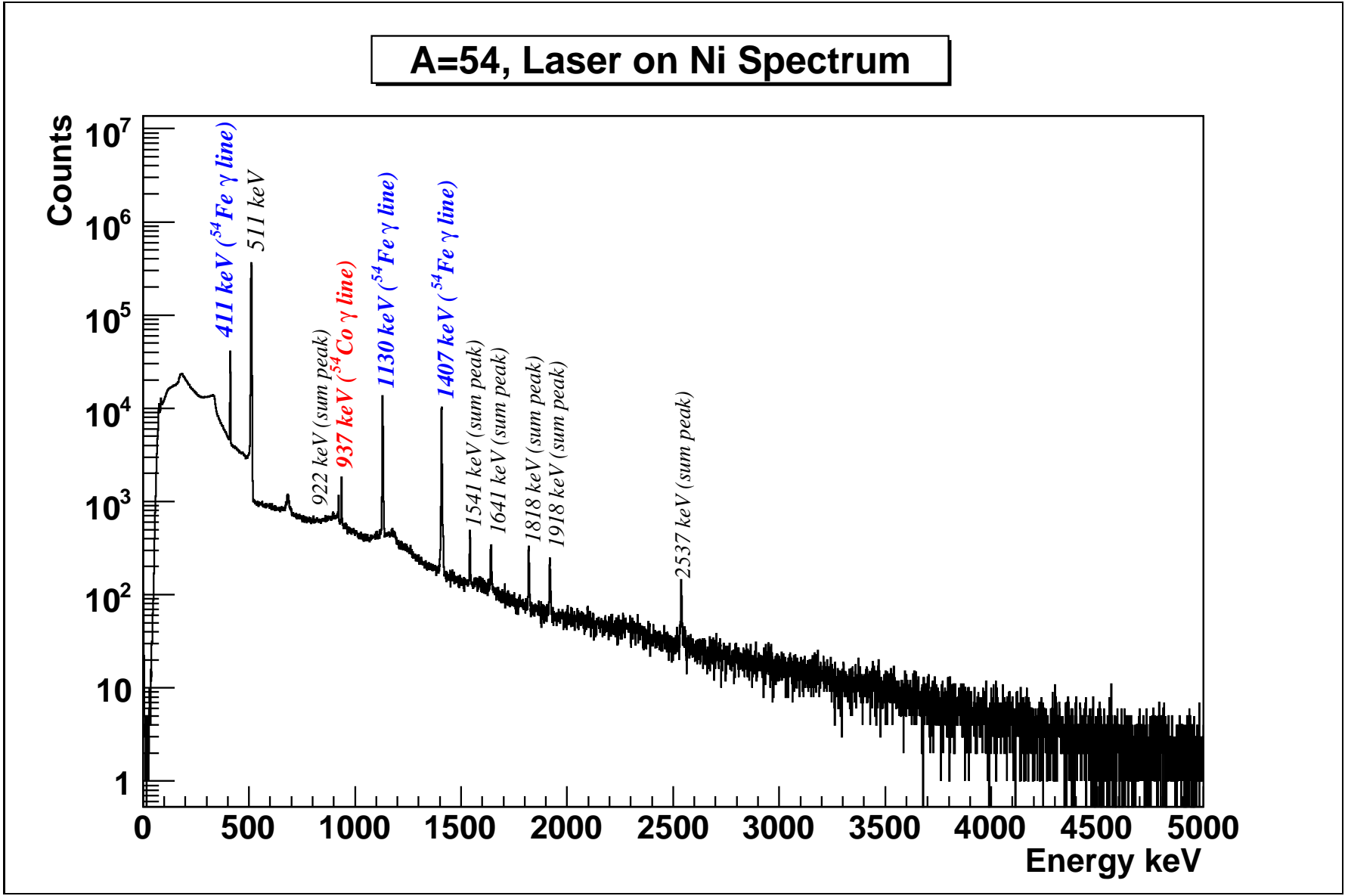}
}\end{center} 
\caption{\emph{A=54 mass, laser on Ni $\beta$-$\gamma$ spectrum}}\label{54Nitotal}
\end{figure}

\begin{figure}
\centering
\fbox{
\includegraphics[angle=0,width=\textwidth]{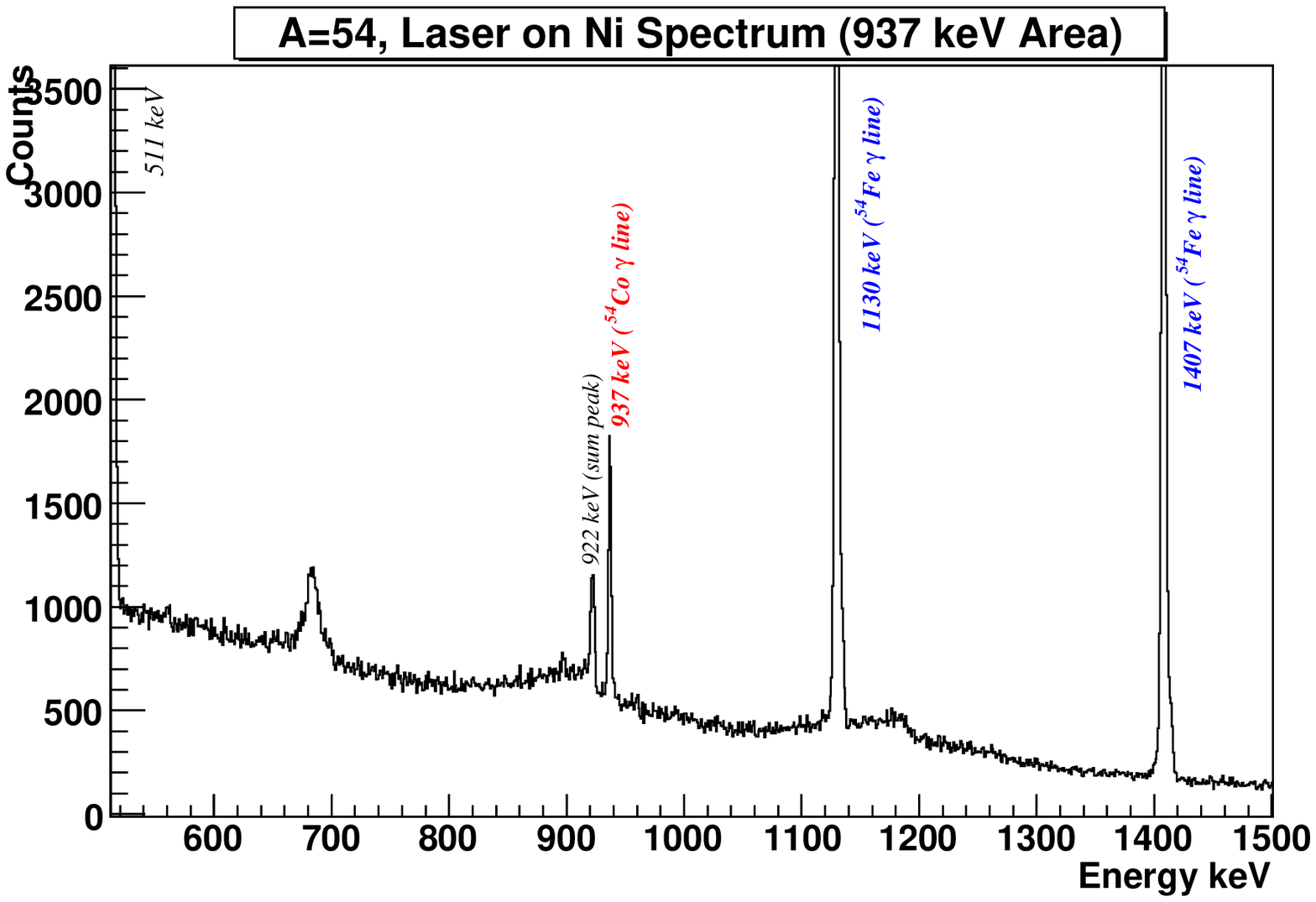}
}
\hspace{3cm}%
\fbox{
\includegraphics[angle=0,width=\textwidth]{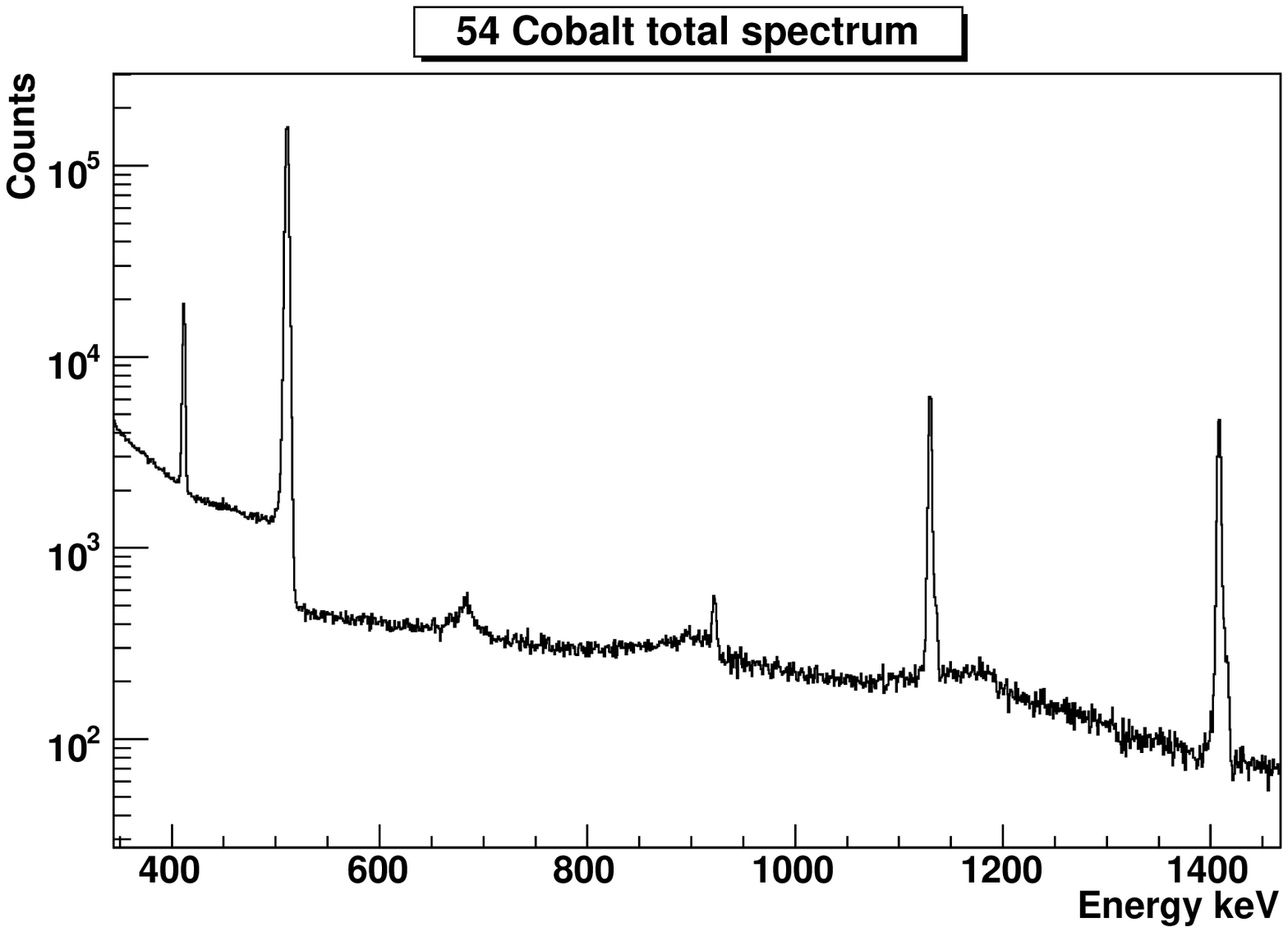}
}
\caption{\emph{Interest zone zoom of the A=54 laser on Ni and laser on Co spectra.}}\label{zoom}
\end{figure}

Figures \ref{54Cototal} and \ref{54Nitotal} represent the total $\gamma$ spectra of mass 54 activity with the laser set on resonace for Ni and Co respectively. Both of them were constructed with  the no-$\beta$ penetration veto condition. In the laser on Ni spectrum we can observe the 937 keV $\gamma$ line known in the decay of $^{54}$Ni. The gamma cascade lines 411 keV, 1130 keV and 1408 keV from the $^{54}$Co$^m$ decay and the 922 keV sum peak (411 keV + 511 keV) are also present. Another peak is observed in the spectra at 683 keV with a measured halflife of $\sim$ 180 ms, but it is not yet identified. See also Figure \ref{zoom} for the low energy part of the spectra. In the laser set on Co spectrum, the same cascade lines and sum peaks are present but not the 937 keV line. 
 
\newpage
\newpage
\section{Halflife Analysis}

\begin{figure}[ht!]
\begin{center}
\framebox{
\includegraphics[angle=0,width=8cm]{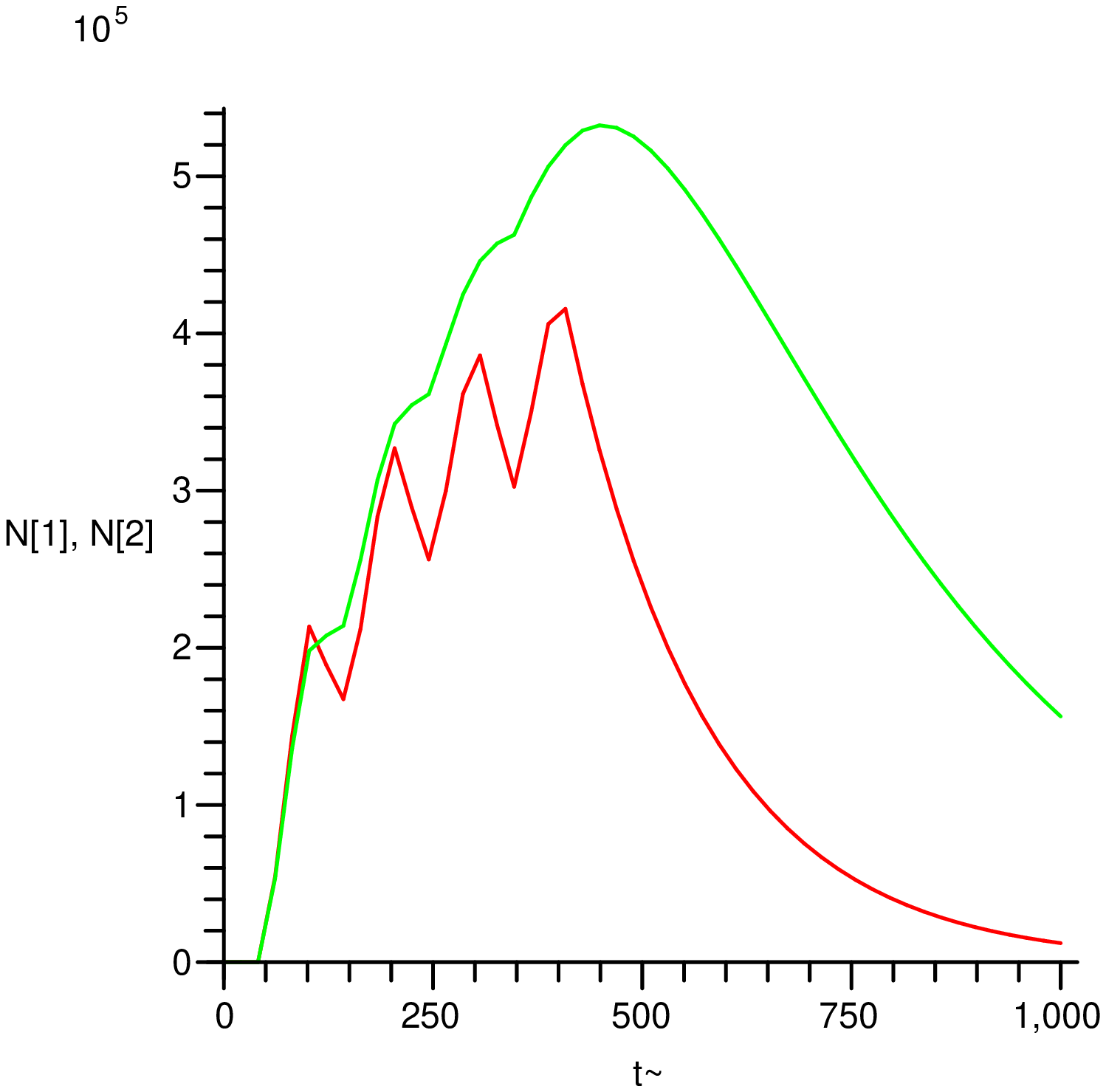}
}\end{center} 
\caption{\emph{ID Curve during a macrocycle time (400 ms implantation and 600 ms decay) solving the differential equations system with Maple10. $^{54}$Ni is shown in red and  $^{54}$Co is shown in green.}}\label{Maple_ID}
\end{figure}

From the previous experiment\cite{Reusen} we know that the $^{54}$Ni  halflife is 106(12) ms, thus if we take into account the macrobeam structure (400 ms of implantation and 600 ms of decay) and the microbeam structure due to the mass separator time behaviour during the implantation, we can model our \emph{Implantation Decay Curve}, simply by solving the decay differential equations.\\
\begin{center}
\begin{equation}
\frac{d}{dt}N_1(t)=P(t)-\lambda_1N_1(t)  \qquad N_1(0)=0
\end{equation}
\begin{equation}
\frac{d}{dt}N_2(t)= \lambda_1N_1(t) - \lambda_2N_2(t) \qquad N_2(0)=0
\end{equation}
\begin{displaymath}
P(t)= \sum_{i=1}^4a(H(t-50(2i-1))-H(t-100i))
\end{displaymath}
\begin{displaymath}
H(t-b)=\left\{ 
\begin{array}{ll}
0 & \textrm{if $t<b$}\\
1 & \textrm{if $t\leq b$}
\end{array}\right.
\end{displaymath}
\end{center}
where $N_1$ is the $^{54}$Ni activity, $N_2$ is the $^{54}$Co activity, $\lambda_1$ and $\lambda_2$ are the  respective decay constants  and $P(t)$ is the implantation rate of $^{54}$Ni, taking into account the microbeam structure. $H(t)$ is the Heaviside step function and $a$ is the constant implantation rate when $P(t)\neq0$.\\
Solving the differential equation with the initial conditions for $N_1(t)$ and $N_2(t)$ using Maple10$^{tm}$, we obtain the \emph{implantation decay curves} of  $^{54}$Ni and $^{54}$Co (Fig.\ref{Maple_ID}).\\

On the other hand, from our experiment we obtain a $\beta$ gated $\gamma$ spectrum, thus  we can reproduce the ID curve looking at the time evolution of the 937 keV peak during one macrocycle. The time difference between each of the spectra (50 ms) was chosen in order to be consistent with the time behaviour of the mass separator. This way we can find the halflife value of $^{54}$Ni by looking at the time behaviour of the integral of the 937 keV $\gamma$ line. 

The analysis of the halflife of  $^{54}$Ni can be deduced from the time behaviour of the area of the 937 keV peak after background substraction in the laser on Ni $\gamma$ spectrum. \\

\begin{figure}[ht!]
\begin{center}
\framebox{

\includegraphics[angle=90,width=\textwidth]{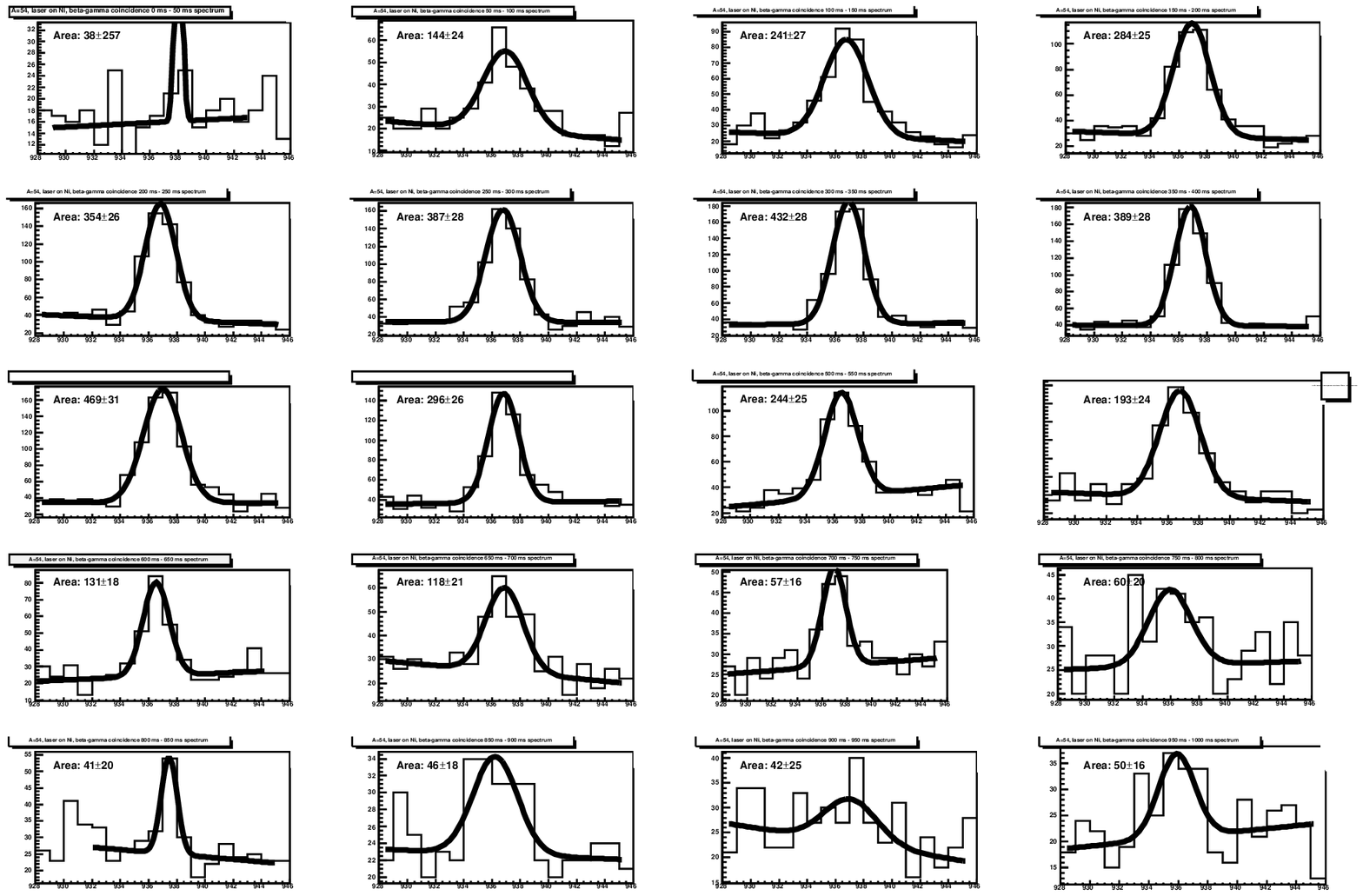}}
  \caption{\emph{'Laser on', Ni $\beta$-$\gamma$ spectra in time. Each of these 20 spectra shows the evolution of the 937 keV $\gamma$ accumulated during 50 ms upto 1000 ms (from left to right and up to bottom), the thick solid line shows the Gaussian fit with a linear background in the near region of the peak. The value of 'Area' written in the spectra is the area of the fitted 937 keV gaussian substracting the long linear background under it.}}\label{Gauss_Fit}
\end{center}
\end{figure}

\begin{figure}[ht!]
\begin{center}
\framebox{
  \includegraphics[angle=0,width=\textwidth]{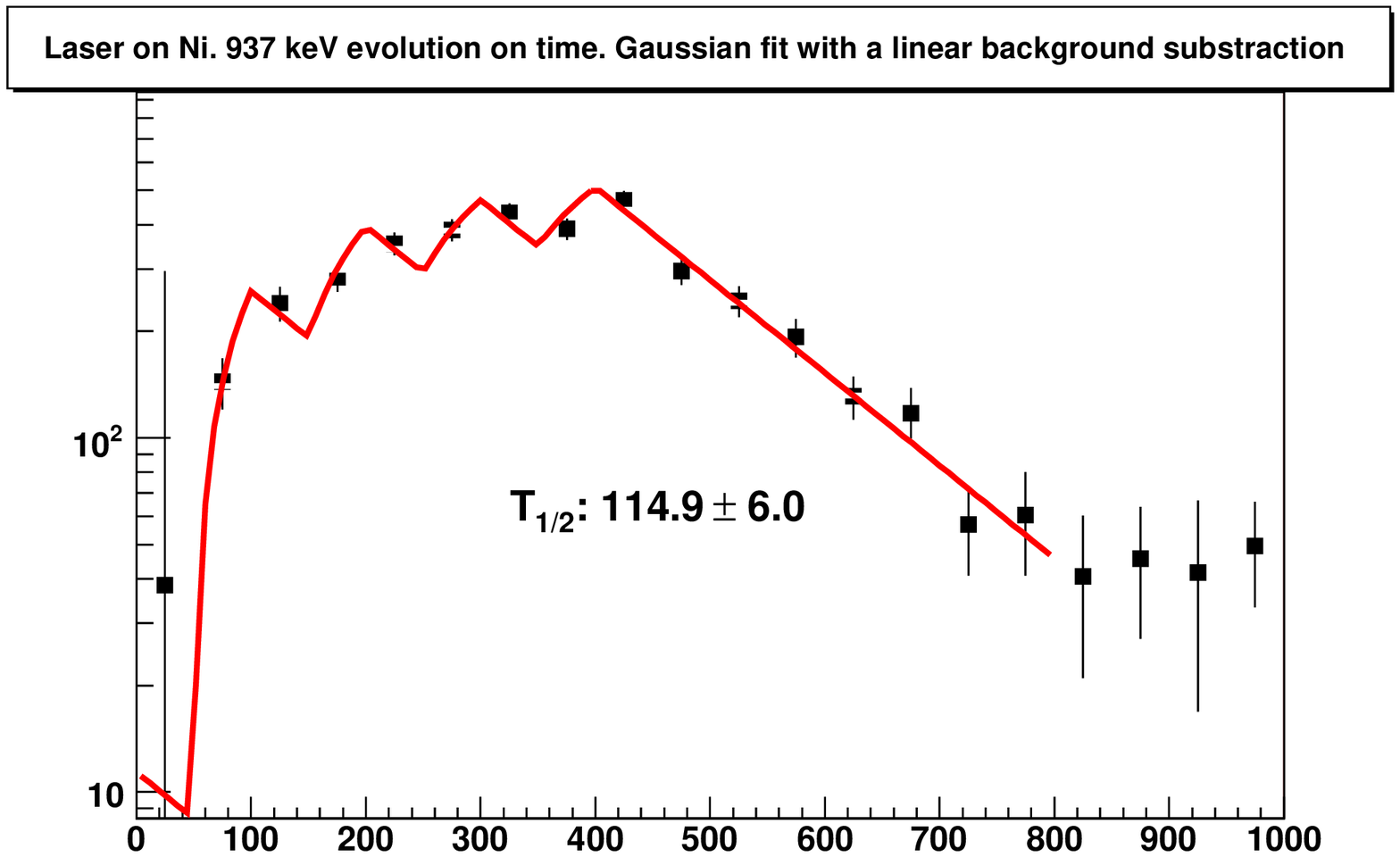}}
  \caption{\emph{Implantation-decay curve of $^{54}$Ni. Each point represents the area of the fitted Gaussian, subtracting a linear background in the near region. }}\label{IDGauss_Fit}
\end{center}
\end{figure}

As a first approach to obtaining the halflife we integrated a  fitted Gaussian with 937 keV centroid, subtracting a linear background close to the Gaussian (centroid $\pm$ twice FWHM region).  Figure \ref{Gauss_Fit} shows the evolution of the 937 keV peak during an ID macrocycle. The last four spectra corresponding to the last 200 ms show that the intensity of the 937 keV line became of the same order as the background. For that reason, the last four points were neglected in the fit of the micro implantation-decay curve (see Figure \ref{IDGauss_Fit}). In this case, the error obtained in fitting a Gaussian for each experimental point in Fig.\ref{IDGauss_Fit} is calculated with the ROOT minimization package MINUIT. \footnote{MINUIT calculates an error matrix with the parameters. This error matrix is the inverse of the second derivative matrix of the best parameter values (the function minimum) and the diagonal elements of the error matrix are the square roots of the individual parameter errors, including the effect of correlation with the other parameters. } The value obtained for the halflife of the $^{54}$Ni in this case was T$_{1/2}$= 114.9 $\pm$ 6.0 ms. The halflife error was obtained from the fit to the implantation-decay curve including the micro structure, again using the ROOT minimization package MINUIT. 

\begin{figure}[ht!]
\begin{center}
\framebox{
\includegraphics[angle=90,width=\textwidth]{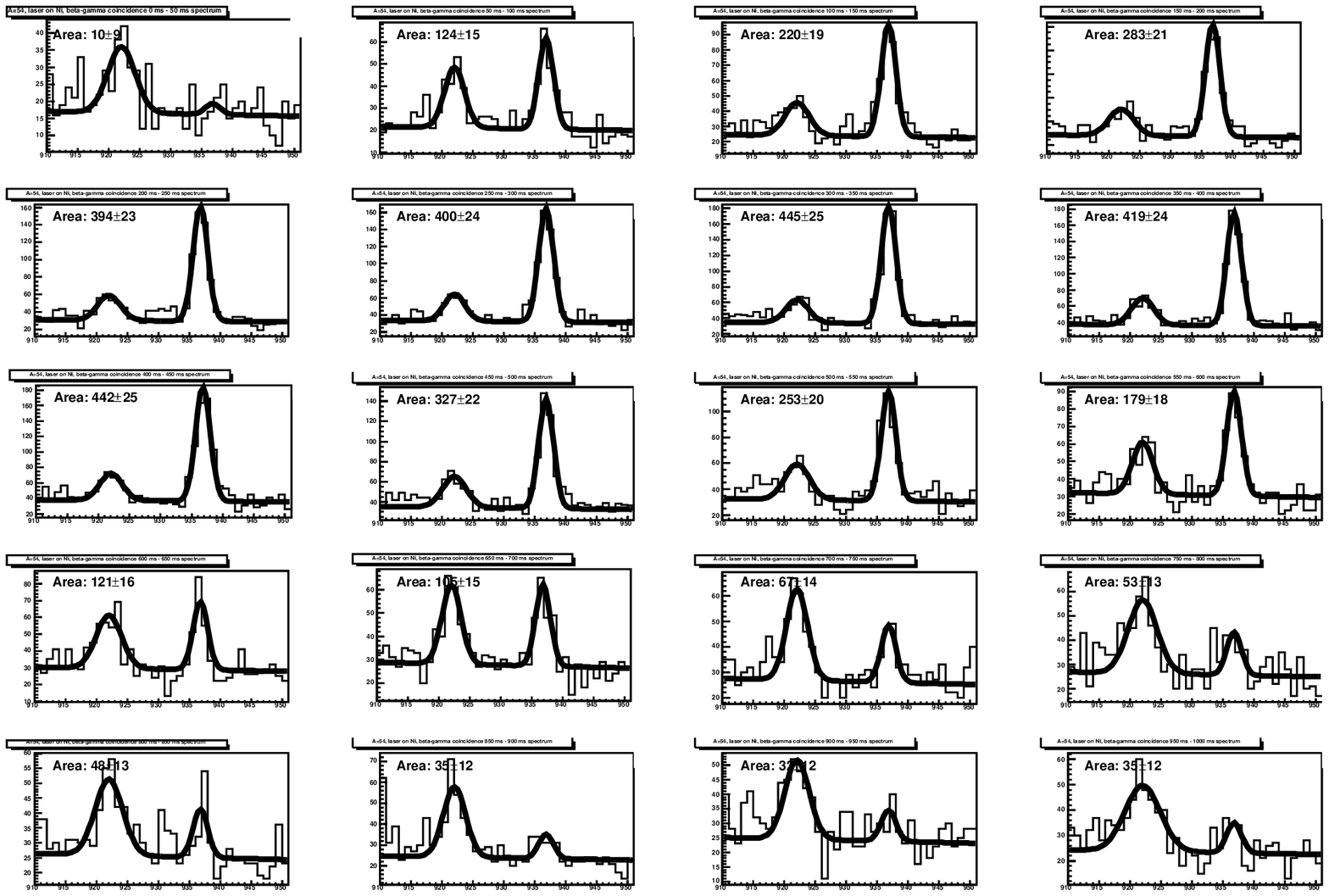}}
  \caption{\emph{'Laser on', Ni $\beta$-$\gamma$ coincidence spectra in time. Each of these 20 spectra shows the evolution of  937 keV $\gamma$ line during a macrocycle every 50 ms, fitting two Gaussians and a long linear background between 800 and 1100 keV. The value of the 'Area' written in the spectra is the area of the fitted 937 keV Gaussian subtracting the long linear background under it.}}\label{Gauss_Fitdobl2}
\end{center}
\end{figure}

Another attempt to deduce the halflife as precisely as possible was made by fitting  two Gaussians, one on the 922 keV sum peak and the other on the 937 peak, the second one with a fixed width (see Figure \ref{Gauss_Fitdobl2}). The value of the width was taken from the width of the Gaussian fit in the total spectra. The so called long background (linear background taken from 800 keV to 1100 keV) was used to subtract the background in the region of interest, in order to reduce background fluctuations in the vicinity of the 937 peak. A reduced error bar is obtained with this method in comparison with the first attempt. Again, the error obtained by fitting a double gaussian for each experimental point in Fig.\ref{IDGauss_Fit} is calculated with the ROOT minimization package MINUIT. However  it is  also clear from this fit that the last points do not behave in the expected way and should therefore be ignored in the $T_{1/2}$ fit. This tells us that due to the lack of statistics (peak indistinguishable from the background)  at the end of the macrocycle, we can eliminate those points in order to fit a proper curve to determine the halflife. The value of the halflife of the $^{54}$Ni for this case was T$_{1/2}$= 113.8 $\pm$ 4.4 ms. The halflife error was obtained from the fit of the implantation-decay curve including the micro structure using again the ROOT minimization package MINUIT.  \\

\begin{figure}[ht!]
\begin{center}
\framebox{
\includegraphics[angle=0,width=\textwidth]{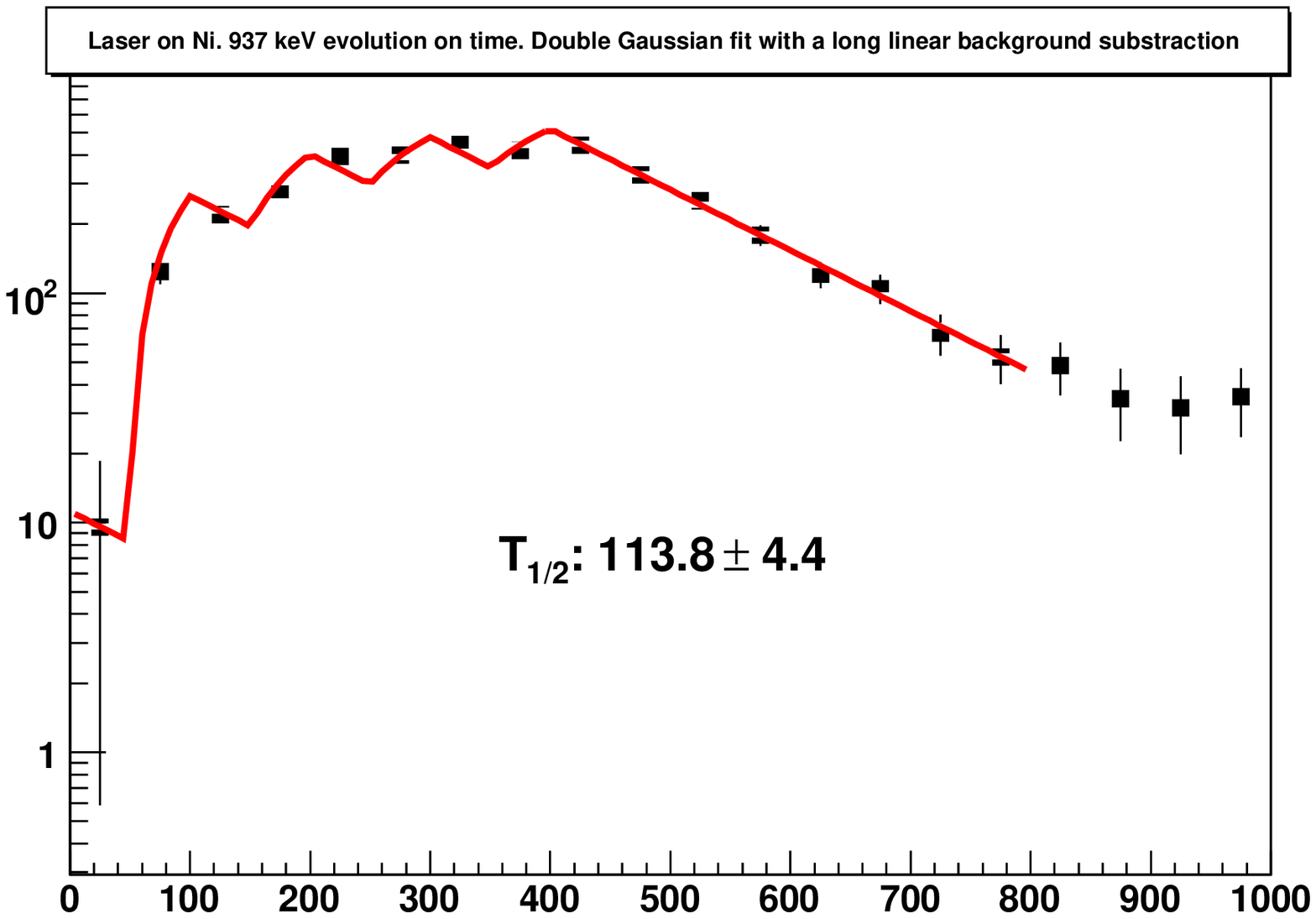}}
  \caption{\emph{Implantation-decay curve of $^{54}$Ni. Each point represent the area of a  Gaussian with fix width and centroid at 937 keV, subtracting a long linear background taken from 800 to 1100 keV, in the region.}}\label{Gauss_fitdob3}
\end{center}
\end{figure}

\begin{figure}[ht!]
\begin{center}
\framebox{
\includegraphics[angle=90,width=\textwidth]{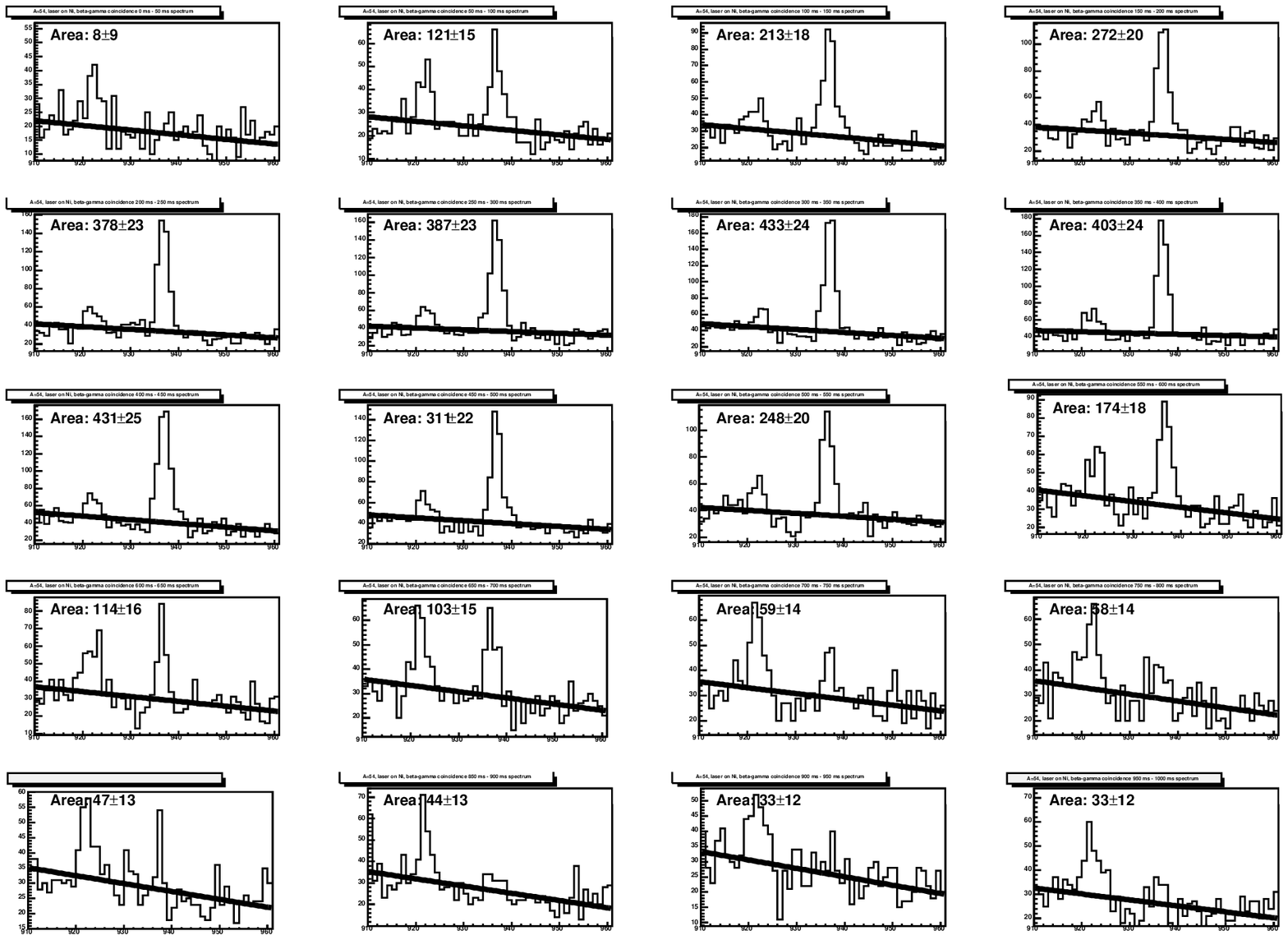}}
  \caption{\emph{$\beta$-$\gamma$ spectra in time. Each spectrum shows the evolution of the 937 keV $\gamma$ line in time steps of 50 ms during a macrocycle, integrating the 937 keV peak and subtracting the long linear background.}}\label{Gauss_Fikdk}
\end{center}
\end{figure}

\begin{figure}[ht!]
\begin{center}
\framebox{
\includegraphics[angle=0,width=\textwidth]{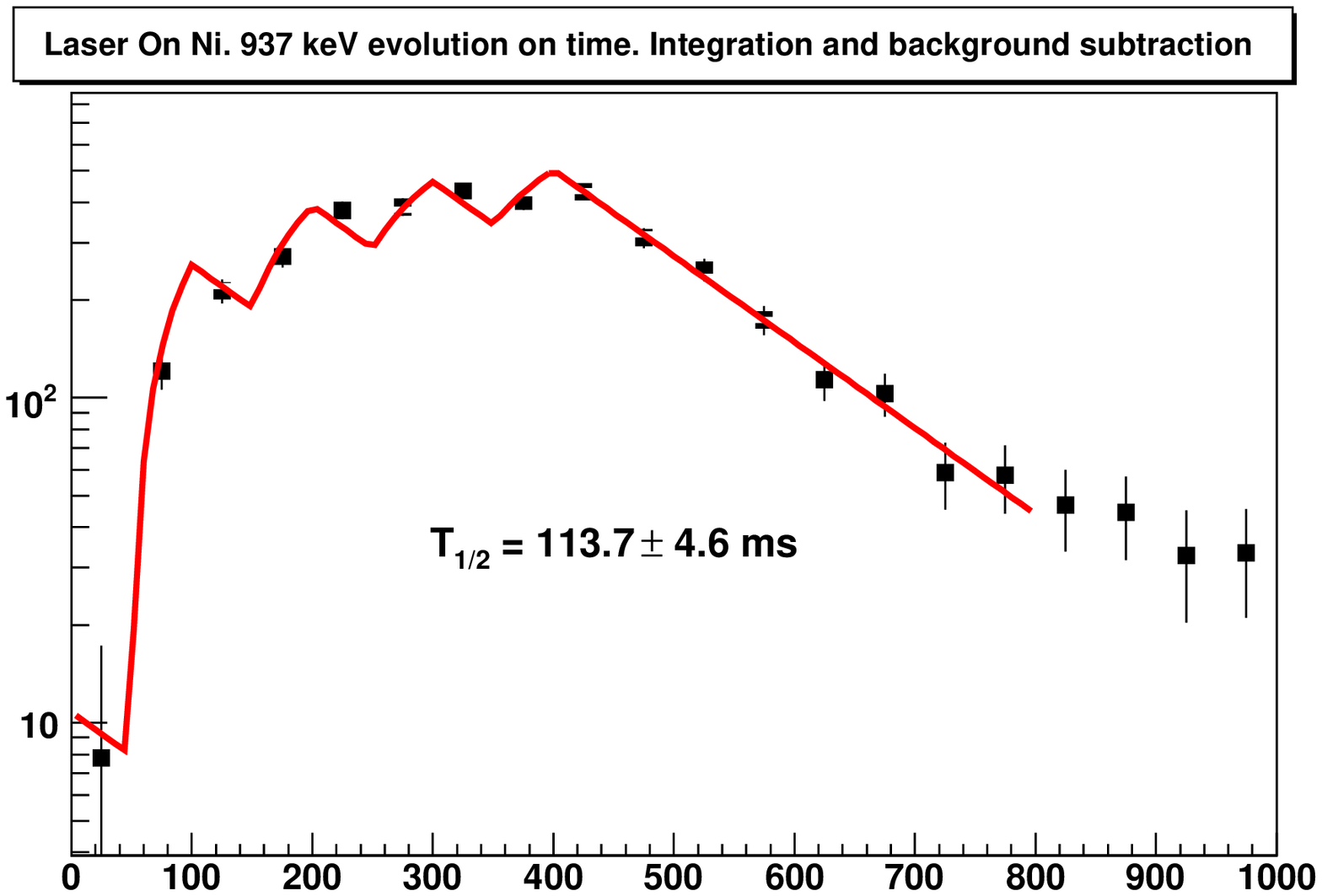}}
  \caption{\emph{Implantation-decay curve of $^{54}$Ni. Each point represents the area of the fitted two Gaussian subtracting a long linear background, taken from 800 to 1100 keV, in the region.}}\label{Gauss_fitdob69}
\end{center}
\end{figure}

A third method to measure the halflife of the $^{54}$Ni was to integrate directly the 937 keV peak and subtract the long linear background (See Figure \ref{Gauss_Fikdk}). The error bar is a statistical error obtained from the integration of the 937 keV peak. If we again neglect the last 200 ms of the macrocycle, the halflife value was T$_{1/2}$= 113.7 $\pm$ 4.6 ms(See Figure \ref{Gauss_fitdob69}). The halflife error was obtained from the fit of the implantation-decay curve including the micro structure using again the ROOT minimization package MINUIT.  \\

In spite of the good agreement shown between these three different halflife analyses, it is worth  mentioning that the estrange behaviour  at the end of the ID curve clearly suggested that some correlation existed between the range of the background chosen and the halflife obtained. Indeed in the 937 keV region, in addition to the 922 keV sum peak, there is the Compton edge of the 1130 keV $\gamma$ peak. This made it very difficult to decide if the background that we were subtracting was the correct one or not.\\
\section{An Attempt to Reproduce the Shape of the Background}
A semi-empirical model for the $\gamma$-ray response of the germanium detectors \cite{Yin} was used to estimate the correct shape of the background for background subtraction under the 937 keV line. In this analysis  was include:
\begin{itemize}
\item[a.-] {\bf The full energy peak}: A Gaussian distribution with variance $B_1^2$ and normalization parameter $A_1$
\begin{equation}\label{ec1}
f_1(E_0,E)= \frac{A_1}{\sqrt{2\pi}B_1} exp \left( \frac{(E-E_0)^2}{2B_1^2}\right)
\end{equation}
\item[b.-] {\bf The Flat continuum}: A constant function when folded with a Gaussian function gives an expression of the form
\begin{equation}
f_2(E_0,E)=A_2 \textrm{\emph{erfc}}\left(\frac{(E-E_0)}{\sqrt{2}B_1}\right)
\end{equation}\label{ec2}
\item[c.-] {\bf The single Compton scattering}: The expression for the energy deposited in a single Compton scattering is well documented in textbooks. In order to account for the roll-off at the Compton edge that is observed in experimental spectra, Y.Jin et al. determine the electron energy function with a complementary error function,
\begin{displaymath}
f_3(E_0,E)=A_3 \left[\left(\frac{E_0}{E_0-E}\right)+\left(\frac{E_0-E}{E_0}\right)-1 +\cos^2\theta \right]\textrm{\emph{erfc}}\left(\frac{(E-B_2)}{\sqrt{2}B_3}\right),\quad E\le E_c
\end{displaymath}
\begin{equation}\label{eq3}
f_3(E_0,E)=0,\qquad E>E_c
\end{equation}
\begin{displaymath}
\cos\theta = 1 + \left(\frac{m_0c^2}{E_0}\right)+\left(\frac{m_0c^2}{E_0-E}\right)
\end{displaymath}
and
\begin{displaymath}
E_c= \frac{E_0}{\left[1+\frac{m_0c^2}{2E_0}\right]}
\end{displaymath}

\item[d.-] {\bf Multiple Compton scattering} Y.Jin et al. have derived expressions for the energy deposition due to two and three Compton scatters. The derivation assumed that the spatial distribution of once-scattered and twice-scattered photons within the detector, are uniform. They give a set of equations for the energy deposition function for the double Compton scattering $f_4(E_0,E)$
\begin{displaymath}
f_4(E_0,E)=0,\qquad   \alpha''>\alpha
\end{displaymath}
\begin{displaymath}
f_4(E_0,E)=A_4\int_{\alpha''}^{\alpha}H(\alpha,\alpha')H(\alpha',\alpha'')d\alpha', \qquad   \frac{\alpha}{1+2\alpha}<\alpha''<\alpha
\end{displaymath}
\begin{displaymath}
f_4(E_0,E)=A_4\int_{\frac{\alpha}{1+2\alpha}}^{\frac{\alpha''}{1-2\alpha''}}H(\alpha,\alpha')H(\alpha',\alpha'')d\alpha', \qquad \frac{\alpha}{1+4\alpha}<\alpha''<\frac{\alpha}{1+2\alpha} 
\end{displaymath}
\begin{equation}
f_4(E_0,E)=0,\qquad   0<\alpha''<\frac{\alpha}{1+4\alpha}
\end{equation}\label{eq4}
where
\begin{displaymath}
H(\alpha,\alpha')=\frac{\alpha'}{\alpha^3}+\left(\frac{1}{\alpha} +\frac{2}{\alpha^2}-\frac{2}{\alpha^3}\right)\left(\frac{1}{\alpha'}\right)+\frac{1}{\alpha^2\alpha'^2}+\frac{2}{\alpha^3}+\frac{1}{\alpha^4},
\end{displaymath}
and
\begin{displaymath}
\alpha=\frac{E_0}{m_0c^2}, \qquad \alpha'=\frac{E'}{m_0c^2},\qquad \alpha''=\frac{E''}{m_0c^2},\qquad E''=E_0-E.
\end{displaymath}
The energy deposition function $f_5(E_0,E)$ for the triple Compton scattering is given by the following set of equation
\begin{displaymath}
f_5(E_0,E)=0,\qquad \alpha''>\alpha,
\end{displaymath}
\begin{displaymath}
f_5(E_0,E)=A_5\int_{\alpha''}^{\alpha}\int_{\alpha''}^{\beta}H(\alpha,\beta)H(\beta,\gamma)H(\gamma,\alpha'')d\gamma d\beta,\qquad \frac{\alpha}{1+2\alpha}<\alpha''<\alpha
\end{displaymath}
\begin{eqnarray*}
 \lefteqn{f_5(E_0,E)=A_5 \int_{\frac{\alpha}{1+2\alpha}}^{\frac{\alpha''}{1-2\alpha''}} \int_{\alpha}^{\beta}H(\alpha,\beta)H(\beta,\gamma)H(\gamma,\alpha'')d\gamma d\beta \quad+}\\
 & & A_5\int_{\frac{\alpha''}{1-2\alpha''}}^{\alpha}\int_{\frac{\beta}{1+2\beta}}^{\frac{\alpha''}{1-2\alpha''}}H(\alpha,\beta)H(\beta,\gamma)H(\gamma,\alpha'')d\gamma d\beta , \qquad \frac{\alpha}{1+4\alpha}<\alpha''<\frac{\alpha}{1+2\alpha}
\end{eqnarray*}
\begin{displaymath}
f_5(E_0,E)=A_5 \int_{\frac{\alpha}{1+2\alpha}}^{\frac{\alpha''}{1-4\alpha''}}\int_{\frac{\beta}{1+2\beta}}^{\frac{\alpha''}{1-2\alpha''}}H(\alpha,\beta)H(\beta,\gamma)H(\gamma,\alpha'')d\gamma d\beta,\quad \frac{\alpha}{1+6\alpha}<\alpha''<\frac{\alpha}{1+4\alpha}
\end{displaymath}
\begin{equation}
f_8(E_0,E)=0, \qquad 0<\alpha''< \frac{\alpha}{1+6\alpha}
\end{equation}\label{eq5}
All these integrals can be solved analytically. 
\end{itemize}

In this same  paper, Jin gives the  parameters $A_i$ and $B_i$ as a function of the energy.
\begin{displaymath}
A_1 = -0.937680+0.203032\times10^1 E_0^{-0.2}-0.892372E_0^{-0.4} \qquad,
\end{displaymath}

\begin{displaymath}
A_2 = -0.110826\times10^{-1}+\frac{0.911437\times10^{-1}}{\sqrt{E_0}} \qquad,
\end{displaymath}

\begin{displaymath}
A_3 = 0.721355\times 10^{-2}+ \frac{0.185895}{E_0^2} \qquad,
\end{displaymath}

\begin{displaymath}
A_4 = -0.775913\times10^1+0.204322\times10^2E_0^{0.1}-0.124189\times10^{2}E_0^{0.2} \qquad,
\end{displaymath}

\begin{displaymath}
A_5 = -0.343254\times10^{-2}+0.213600\times10^{-1}E_0-0.882892\times10^{-3}E_0^2 \qquad,
\end{displaymath}
\begin{displaymath}
B_2 = -0.156849+0.965065 E_0 \qquad \textrm{, and}
\end{displaymath}
\begin{displaymath}
B_3= 0.443463\times10^{-1}+0.1697733\times10^{-1}E_0 \qquad.
\end{displaymath}

\begin{figure}[ht!]
\begin{center}
\framebox{
\includegraphics[angle=0,width=8cm]{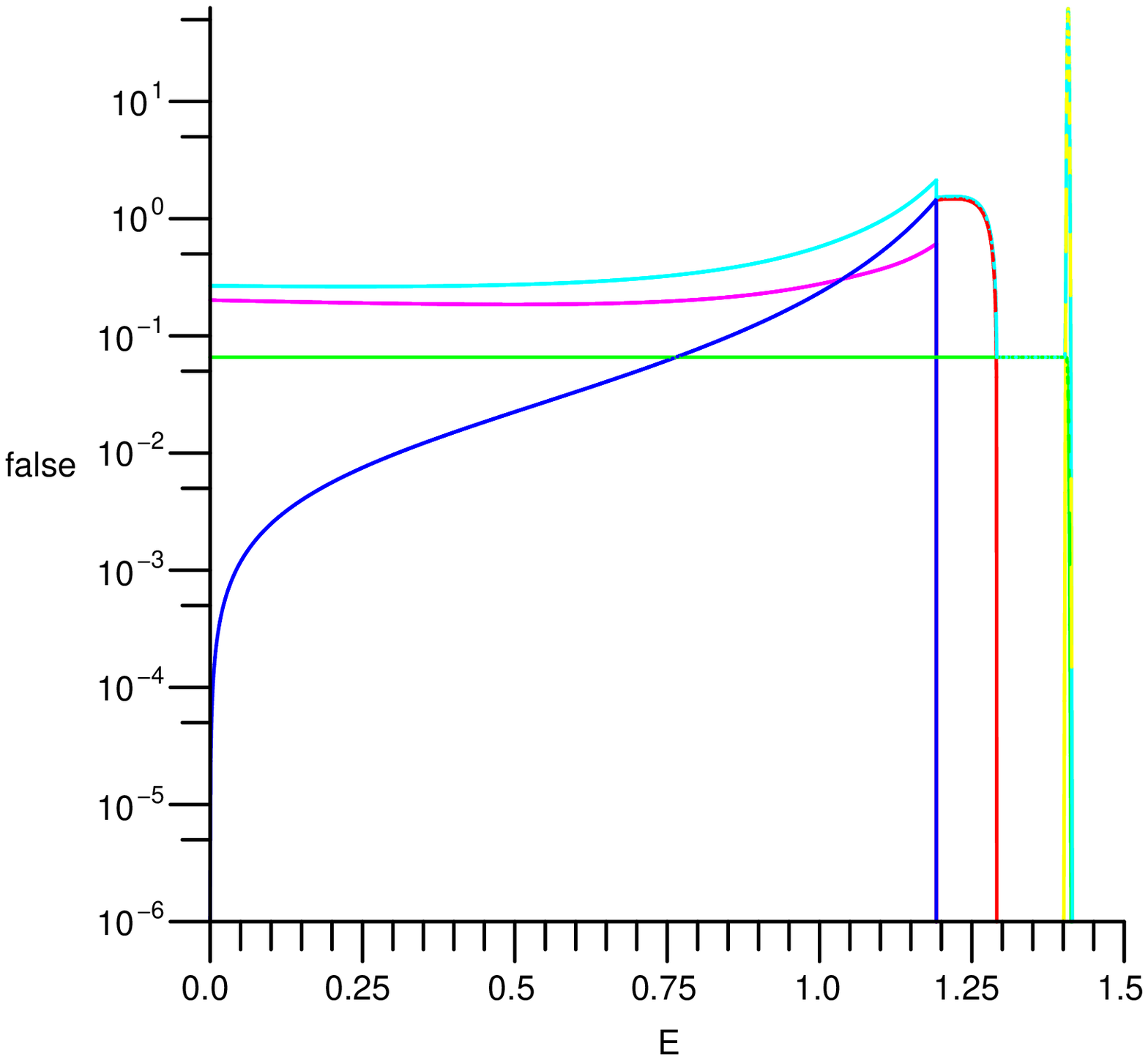}}
  \caption{\emph{Background estimated from Jin's semi empirical method. The pink line represents the single Compton scattering, the yellow line is the Gaussian peak,  the green line is the flat continuum, the red line is the double Compton scattering and the blue line is the triple Compton scattering. The light blue line is the sum of all of them.}}\label{maplegraphs}
\end{center}
\end{figure}

As a first attempt to fit the calculated background we used the 1407 keV peak. This peak was chosen because its Compton edge can be easily distinguished in the $^{54}$Ni and $^{54}$Co spectra. The integrals in (3.7) and (3.8),  were solved analytically with MAPLE10$^{tm}$ and its solutions with the full energy peak, single Compton scattering and the flat continuum are presented in Figure \ref{maplegraphs}.  

\begin{figure}[ht!]
\begin{center}
\framebox{
\includegraphics[angle=0,width=8cm]{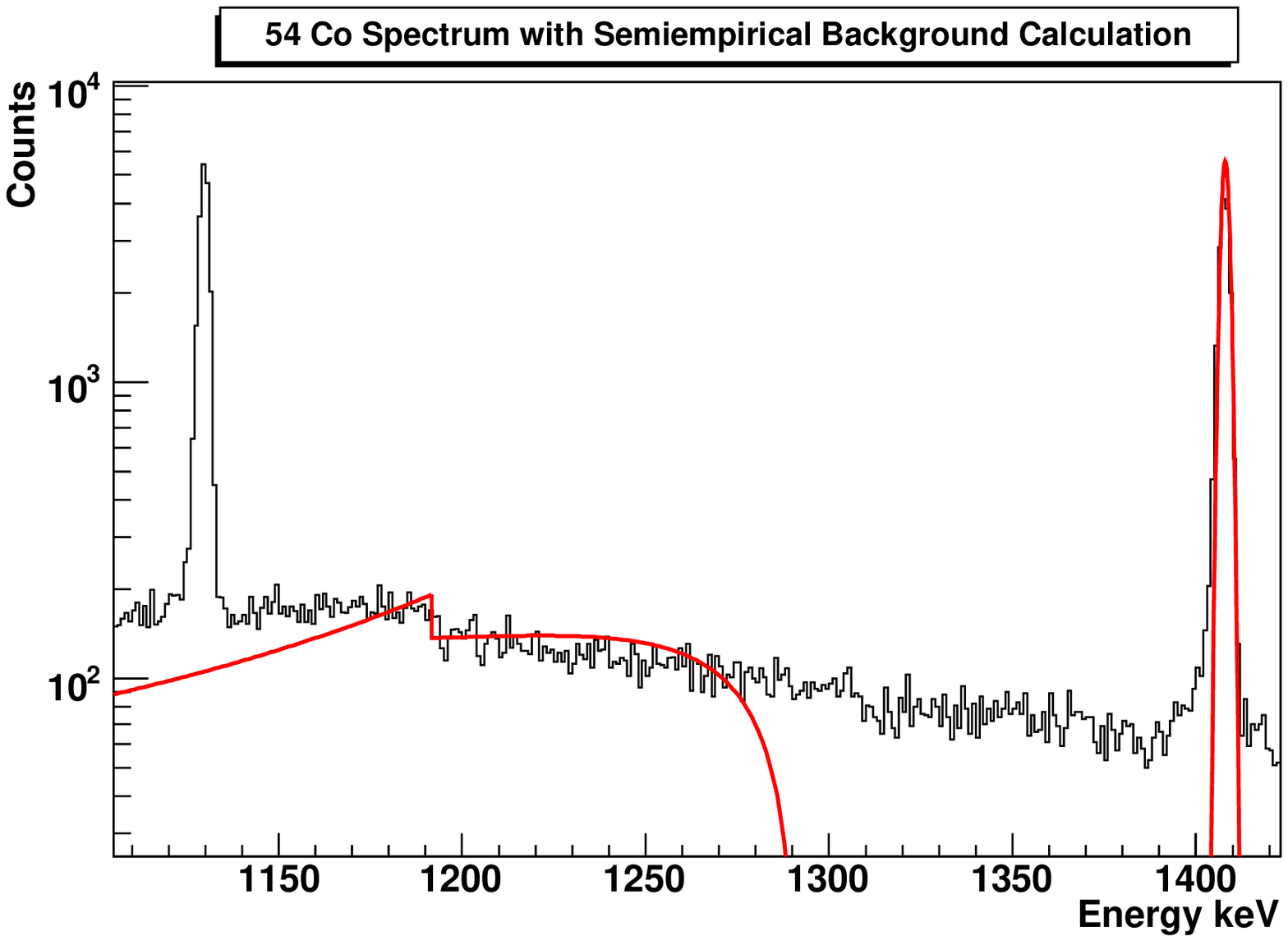}}
  \caption{\emph{$^{54}$Co Spectrum is shown in black and the semiempirical background calculation is shown in red.}}\label{Jinbkg}
\end{center}
\end{figure}

In Figure \ref{Jinbkg} is shown the $^{54}$Co total spectrum with the calculated background. A  correction factor was applied to this calculated  background in order to equal the amplitude of the full energy peak of 1407 keV with the measured spectrum. A good agreement between the theoretical background and the real data is presented, but no further analysis  has been done up to the time of writing.

\section{High Energy Gamow-Teller states}
\begin{figure}[ht!]
\begin{center}
\framebox{
\includegraphics[angle=0,width=\textwidth]{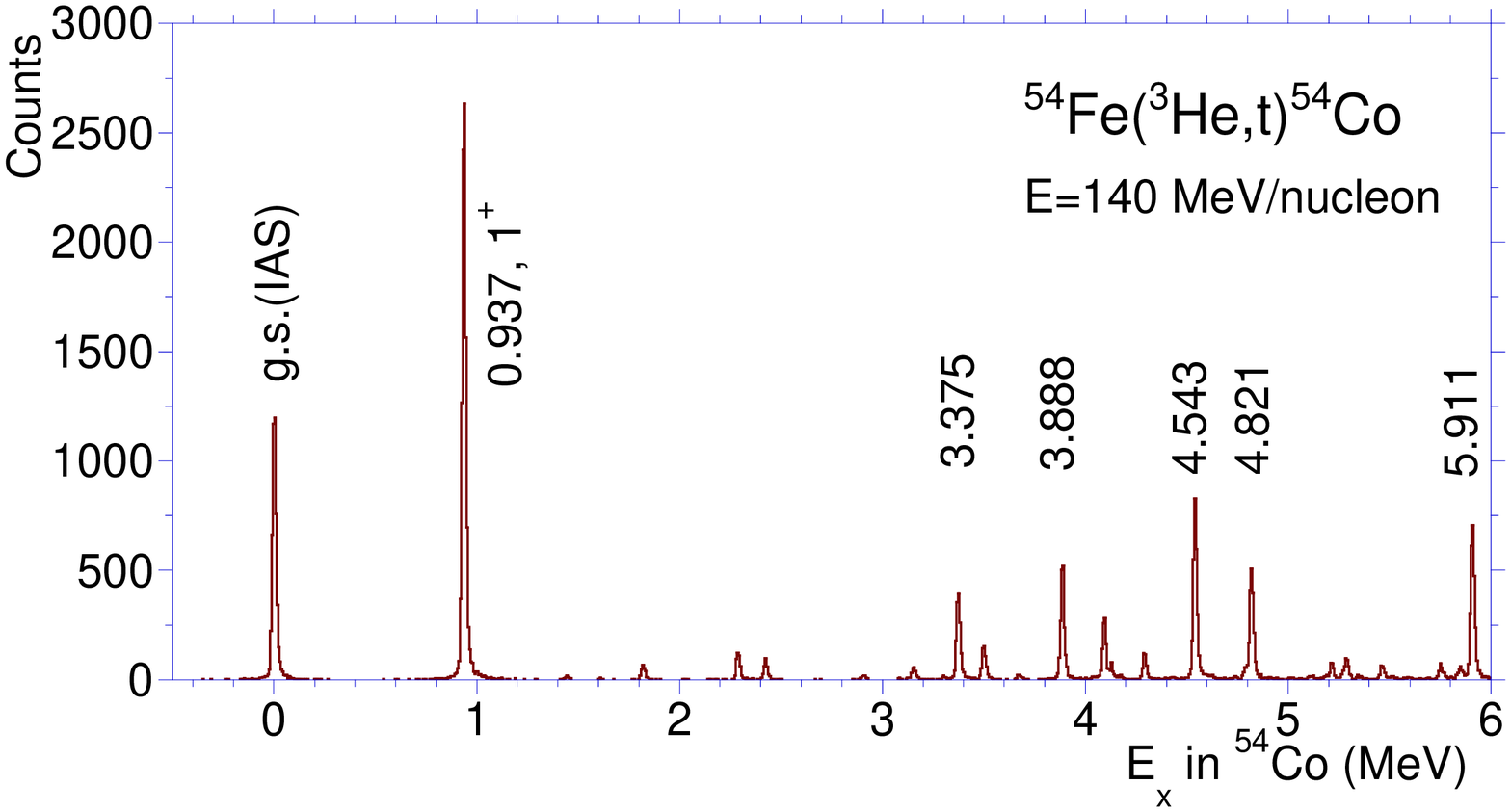}}
  \caption{\emph{Energy spectum of the $^{54}$Fe($^3$He,t)$^{54}$Co charge exchange reaction.}}\label{chex}
\end{center}
\end{figure}

In  Figure \ref{chex} we present the energy spectra obtained in the $^{54}$Fe($^3$He,t)$^{54}$Co charge exchange reaction \cite{privYoshi}. In addition to the $0^+\rightarrow 0^+$ Fermi transition and the first $0^+\rightarrow 1^+$ Gamow-Teller transition, other high energy  states are populated in the reaction. Because in the $\beta$ decay case we have a $Q$-value window up to $\sim$8.8 MeV we also expected to populate some of these states. Because of the expected drop in $\beta$-branching due to the $f$ factor and the drop of the $\gamma$-ray detection efficiency at high $\gamma$-energies, we expect to see the 3375 keV peak more easily than the other higher lying states. We therefore searched for the 3375 keV peak.

\begin{figure}[ht!]
\begin{center}
\framebox{
\includegraphics[angle=0,width=\textwidth]{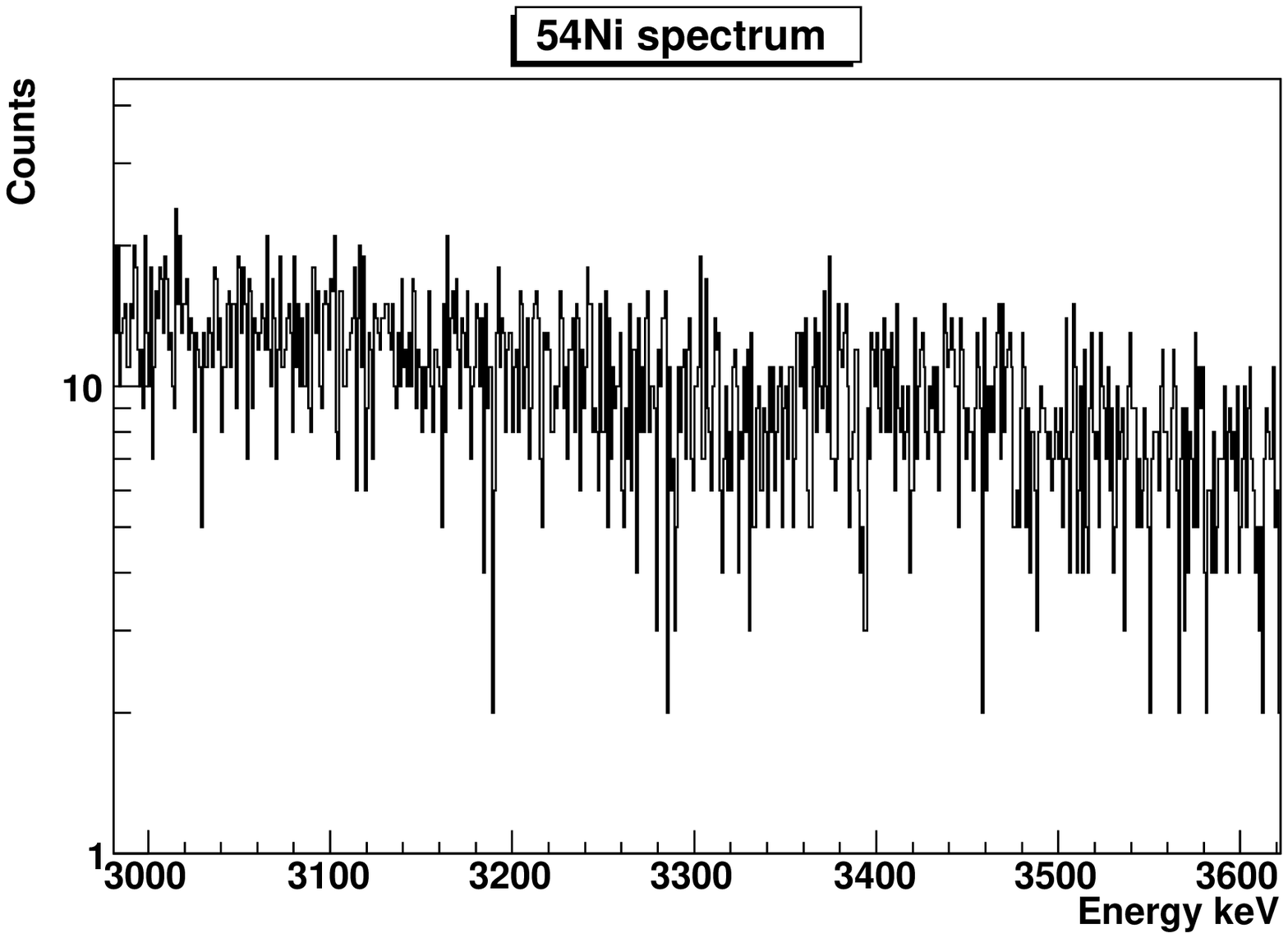}}
  \caption{\emph{$^54$Ni spectrum in the 3375 keV vicinity.}}\label{54Nihigh}
\end{center}
\end{figure}
In the Figure \ref{54Nihigh}  the 3375 keV region is plotted, but no $\gamma$ lines are distinguishable from the background. 
\newpage
\section{Final $T_{1/2}$ and Absolute Gamow-Teller calculation}
With the three values obtained from the behaviour of the 937 keV $\gamma$ line the  value of the $^{54}$Ni halflife is calculated as the weighted mean \footnote{This is a preliminar estimation of the halflife, and certainly is not the right way to do it because this three values are not calculated from independent measurements.} which value is $T_{1/2}=114 \pm 3$ ms. With this value we can estimate the absolute Gamow-Teller strength, using  equation \ref{yfuji}. 
\begin{displaymath}
\frac{1}{T_{1/2}}= \frac{B(F)(1-\delta_c)}{K\sigma^F}\left( \sigma^F f_F + \frac{\lambda^2}{R^2}  \sum_{i=GT}\sigma_i^{GT}f_i \right)
\end{displaymath}
where $T_{1/2}=0.114(3)$ s is the measured halflife, $B(F)=2$ is the $\vert N-Z \vert$ value, $(1-\delta_c)=0.995$ is the Coulomb correction factor, $\sigma^F=4905\pm245$ is the number of counts in the $0^+\rightarrow 0^+$ transition in the charge exchange reaction (see Figure \ref{chex} ), $f_F=22107\pm 746$ and $\sum_{i=GT} \sigma_i^{GT}f_i=(130.9\pm8.1)\times10^6$, $K=6144.4\pm 1.6$ where the $f$ values are given by the $f$-function calculation, and  $\lambda^2=1.603\pm0.007$. With these values we can calculate the $R^2$ value which is related to the $B(GT)$ as is shown in Eq. (1.24).
\begin{eqnarray*}
\lefteqn{
\frac{1}{0.114 \pm 0.003} = \frac{2\times0.995}{(6144.4 \pm 1.6) \times(4905\pm 245)} \quad\times}\\
& &
\times\left[(1.0844\pm 0.0908) \times10^8+\frac{1.603\pm 0.007}{R^2}\times(130.9\pm 8.1)\times10^6\right]
\end{eqnarray*}
 
 \begin{displaymath}
 R^2=8.56 \pm 0.73 
 \end{displaymath}
 \begin{equation}
 B(GT)_{(T_{1/2}=114\pm3)}=0.48 \pm 0.08 \quad.
 \end{equation}
 
The calculated  absolute Gamow-Teller strength  is  $B(GT)=0.48 \pm 0.08$.

\chapter{Conclusions and Future}

The change exchange reactions and the $\beta$-decay process are complementary ways to study the Gamow-Teller strength. The advantage of the charge exchange reactions are that they are not limited by the $Q$-value as the $\beta$-decay is. But the big advantage of the $\beta$-decay is that the measurement of the Gamow-Teller strength to one particular state is sufficient to normalize the Gamow-Teller strength obtained by the charge exchange reactions. \\
In this report a $\beta$-decay experiment was described and analyzed. The total halflife of the $^{54}$Ni was obtained from the $\gamma$ spectra giving a tentative value of $T_{1/2}=114\pm3$ ms. In order to improve and facilitate the halflife analysis, a semi-empirical model for the $\gamma$-ray response function of a Ge detector was used to obtain the background shape including multiple Compton scattering and the Compton edge. This phase of the analysis is still under development but we already obtained satisfactory results if we compare the shape of the estimated background with the real shape  obtained in our experiment. \\
In the near future we will start with the  analysis of the $\beta$s in order to obtain the branching ratio between the super-allowed Fermi and the first Gamow-Teller states.\\
A new experiment to measure the decay of $^{54}$Ni has been performed very recently at GSI (Darmstadt, Germany) and the on-line analysis showed very promising results. The main advantage of this facility is the high rate of production of $^{54}$Ni in the Fragment Separator ($\sim$100 $^{54}$Ni ions per second, in comparison to $\sim$ 3600 total counts at LISOL). For that reason we hope to see the high energy Gamow-Teller state in the $\beta$-decay, as well. 
\renewcommand{\chaptermark}[1]{\markboth{Ap\'endice\ \thechapter}{}} 
\renewcommand{\sectionmark}[1]{\markright{\thesection\ #1}}
\pagestyle{empty}
\bibliographystyle{alpha}

\end{document}